\documentclass[fleqn,usenatbib]{mnras}
\usepackage{newtxtext,newtxmath}
\usepackage[T1]{fontenc}
\DeclareRobustCommand{\VAN}[3]{#2}
\let\VANthebibliography\thebibliography
\def\thebibliography{\DeclareRobustCommand{\VAN}[3]{##3}\VANthebibliography}
\usepackage{graphicx}
\usepackage{amsmath}

\defcitealias{Birrer2020}{TDCOSMO IV}

\title[$H_0$ and strong lensing selection biases]{Correcting for Selection Biases in the Determination of the Hubble Constant from Time-Delay Cosmography}

\author[Tian Li et al.]{
Tian Li$^{1}$\thanks{E-mail: tian.li@port.ac.uk},
Thomas E. Collett$^{1}$,
Philip J. Marshall$^{2}$,
Sydney Erickson$^{2}$,
Wolfgang Enzi$^{1}$,
\newauthor
Lindsay Oldham$^{1}$,
Daniel Ballard$^{1,3}$.
\\
$^{1}$ Institute of Cosmology and Gravitation, University of Portsmouth, Burnaby Rd, Portsmouth PO1 3FX, UK\\
$^{2}$ Kavli Institute for Particle Astrophysics and Cosmology, P.O. Box 20450, MS29, Stanford, CA 94309, U.S.A.\\
$^{3}$ Sydney Institute for Astronomy, School of Physics, University of Sydney, NSW 2006, Australia
}
\date{Accepted XXX. Received YYY; in original form ZZZ}

\pubyear{\the\year{}}

\begin{document}
\label{firstpage}
\pagerange{\pageref{firstpage}--\pageref{lastpage}}
\maketitle

\begin{abstract}

The time delay between multiple images of strongly lensed quasars has been used to infer the Hubble constant. The primary systematic uncertainty for time-delay cosmography is the mass-sheet transform (MST), which preserves the lensing observables while altering the inferred $H_0$. The TDCOSMO collaboration used velocity dispersion measurements of lensed quasars and lensed galaxies to infer that mass sheets are present, which decrease the inferred $H_0$ by 8$\%$. Here, we test the assumption that the density profiles of galaxy-galaxy and galaxy-quasar lenses are the same. We use a composite star-plus-dark-matter mass profile for the parent deflector population and model the selection function for galaxy-galaxy and galaxy-quasar lenses. We find that a power-law density profile with an MST is a good approximation to a two-component mass profile around the Einstein radius, but we find that galaxy-galaxy lenses have systematically higher mass-sheet components than galaxy-quasar lenses. For individual systems, $\lambda_\mathrm{int}$ correlates with the ratio of the half-light radius and Einstein radius of the lens. By propagating these results through the TDCOSMO hierarchical inference code, we find that $H_0$ is lowered by a further $\sim$3\%. Using a more recent measurement of velocity dispersions and our fiducial model for selection biases, we infer $H_0 = 66\pm4 \ \mathrm{(stat)} \pm 1 \ \mathrm{(model \ sys)} \pm 2 \ \mathrm{(measurement \ sys)} \ \mathrm{km} \ \mathrm{s}^{-1} \ \mathrm{Mpc}^{-1}$ for the TDCOSMO plus SLACS dataset. The first residual systematic error is due to plausible alternative choices in modeling the selection function, and the second is an estimate of the remaining systematic error in the measurement of velocity dispersions for SLACS lenses. Accurate time-delay cosmography requires precise velocity dispersion measurements and accurate calibration of selection biases.

\end{abstract}



\section{Introduction}

The Universe's current expansion rate is called the Hubble constant, $H_0$. $H_0$ is fundamental in setting the age and scale of our Universe, but despite nearly a century of research the value of $H_0$ is not definitively known. Local Universe measurements ( $z < 1$ ) by the SH0ES collaboration combine, parallax, Cepheid variables and Type Ia supernovae to calculate distances in the Hubble flow, finding $H_0 = 73.0 \pm 1.0\,\mathrm{km\,s}^{-1}\,\mathrm{Mpc}^{-1}$
\citep{Riess2022}. Another method infers $H_0$ by extrapolating from observations of the Cosmic Microwave Background (CMB) radiation in the early universe (z $\approx$ 1100), which provides a lower value of $H_0=67.4 \pm 0.5\,\mathrm{km\,s}^{-1}\,\mathrm{Mpc}^{-1}$ when interpreted within the framework of the standard $\Lambda$CDM model \citep{2020A&A...641A...6P}. These two measurements of $H_0$ disagree by $9\%$, and this disagreement is colloquially referred to as the Hubble tension. This persistent disagreement suggests either there are systematic errors that are poorly understood or new physics beyond the standard cosmological model \citep{Verde:2019ivm}.

Multiple independent methods have been developed to determine the cause of the ``Hubble tension.'' Spectroscopic cosmological surveys of galaxies measure the large-scale structure that originates from the density fluctuations of the early universe. The Baryon Acoustic Oscillations (BAO) from Dark Energy Spectroscopic Instrument (DESI) gives $H_0=68.52\,\pm\,0.62\,\mathrm{km\,s}^{-1}\,\mathrm{Mpc}^{-1}$ when combining with a baryon density prior from Big Bang Nucleosynthesis (BBN) \citep{DESI:2024mwx}. Another BAO measurement from Dark Energy Survey (DES) gives $H_0=67.4 \pm 1.2 \,\mathrm{km} \mathrm{s}^{-1} \mathrm{Mpc}^{-1}$. Under the same BBN prior, the ``Full Shape'' analysis of the galaxy power spectrum from the Baryon Oscillation Spectroscopic Survey (BOSS) yields $H_0=69.6_{-1.3}^{+1.1}\,\mathrm{kms}^{-1}\,\mathrm{Mpc}^{-1}$ \citep{Philcox:2021kcw}. The above early universe observations are in agreement with the lower value of H0.

Late-universe probes yield a wider range of $H_0$ values. Modifying the calibration steps of the local distance ladder shifts the inferred $H_0$. The Carnegie-Chicago Hubble Project, for instance, uses the tip of the red giant branch (TRGB) instead of Cepheids, resulting in a measurement of $H_0 = 69.6\pm1.7\,\mathrm{km} \mathrm{s}^{-1}  \mathrm{Mpc}^{-1}$ \citep{Freedman:2019jwv}. When calibrating distance ladder using Mira variable stars, $H_0$ is determined to be $73.3 \pm 4.0\,\mathrm{km} \mathrm{s}^{-1} \mathrm{Mpc}^{-1}$ \citep{Huang:2019yhh}. Independently of the distance ladder, megamasers provide a direct geometrical measurement of $H_0$. The Megamaser Cosmology Project (MCP) measured the distances of several nearby galaxies, yielding $H_0 = 73.9\pm 3.0\,\mathrm{km} \mathrm{s}^{-1} \mathrm{Mpc}^{-1}$ \citep{Pesce:2020xfe}.

The Hubble constant can also be measured by observing the time delay in the arrival time between different images in a gravitational lensing system \citep{1964MNRAS.128..307R}. The H0LiCOW collaboration used 6 lensed quasars to constrain $H_0$ to 2.4$\%$ and found $H_0=73.3_{-1.8}^{+1.7} \mathrm{~km} \mathrm{~s}^{-1} \mathrm{Mpc}^{-1}$ using either a power-law profile or a star-plus-dark-matter profile. Adding the blind measurement by \cite{Shajib2020} further increased the precision to approximately 2$\%$ \citep{Millon:2019slk}.

The primary source of uncertainty in measuring $H_0$ with gravitational lenses is the mass-sheet transform (MST). By adding a constant mass sheet and rescaling the convergence, the MST changes the shape of the galaxy mass profile while keeping all the lensing observables unchanged except for time delays and absolute magnifications \citep{1985ApJ...289L...1F}. \citet{Schneider:2013sxa} showed that fitting a power law density profile to lensing data can result in up to a  $\sim$20$\%$ bias in $H_0$ if the true profile is a composite of stars and dark matter. \citep{Kochanek:2019ruu, Kochanek:2020crs} found similar results (although \citet{2020A&A...639A.101M} found consistent result by modeling with a power-law profile and NFW+stars profile). Whilst measurements of velocity dispersions can help to break the MST,  kinematics measurements typically lack the signal-to-noise needed and interpreting them introduces further potential systematics.

To overcome the MST, \cite{Birrer2020} (hereafter TDCOSMO IV) fit a more flexible density profile to the H0liCOW dataset and quantified the MST based on kinematic data of galaxies lensing quasars and 33 SLACS galaxy-galaxy lenses. When using kinematic data from lensed quasar only, the H0 after quantifying MST is $H_0=74.5_{-6.1}^{+5.6} \mathrm{~km} \mathrm{~s}^{-1} \mathrm{Mpc}^{-1}$. Combining with the MST from 33 SLACS galaxy-galaxy lenses, they found $H_0=67.4_{-3.2}^{+4.1} \mathrm{~km} \mathrm{~s}^{-1} \mathrm{Mpc}^{-1}$. The key assumption made in this paper is that the population of SLACS deflector galaxies have the same underlying mass profiles and kinematic properties as lensed quasars. However, this is unlikely to be true: \citet{Collett:2016muz} found that we should expect the lensed quasar population to be biased towards shallow density profiles. \citet{Xu:2015dra} showed that $\gamma$ and $\lambda_{\text {int }}$ are correlated for elliptical galaxies in the Illustris simulation, so any difference in the selection function of lensed quasars and galaxy-galaxy lenses is likely to propagate into biases on $H_0$. 


\citet{Gomer2022} demonstrate that using lenses with a smaller Einstein radius relative to their effective radius can improve the accuracy of the \(H_0\) measurement, although they did not account for the selection effect between the observed SLACS and TDCOSMO samples. In \citet{Sonnenfeld2023}, the authors examined the selection effect by modeling the stellar mass distribution with a de Vaucouleurs profile and the dark matter distribution with a generalized Navarro–Frenk–White (gNFW) profile for lenses with extended sources and those producing lensed quasars. They found that contrary to \citetalias{Birrer2020}, the intrinsic parameter \(\lambda_{int}\) for quad lenses is approximately 2\% smaller than that for lenses with extended sources. 

In this work, we aim to understand the differences in lens mass properties between galaxy-galaxy and galaxy-quasar lenses, and the implications of these differences when combining these samples for an inference of $H_0$ with different kinds of mass models. We simulate lens galaxies using a two-component model with the Sersic \cite{1963BAAA....6...41S} and Navarro–Frenk–White (NFW) \citep{1997ApJ...490..493N} profiles of the stars and dark matter halos respectively. We perform mock observations to select SLACS-like and TDCOSMO-like lens systems. We then compare the population differences of $\lambda_\mathrm{int}$. Finally, we propagate these differences through the hierarchial inference codes of \citetalias{Birrer2020} to recalibrate $H_0$. The purpose of our work is to study the impact of selection effect between different lens populations, and the choice of different mass models, we did not blind our analysis.

The paper is structured as follows: Section \ref{sec:theory} describes the principles of time-delay cosmography and the mass-sheet transform. Section \ref{sec:simulation} explains the construction of our simulation sample and the methodology for computing $\lambda_\mathrm{int}$. In Section \ref{sec:result}, we present the parameters of different populations. We then adapt the methodology described in \citetalias{Birrer2020} and present the $\lambda_\mathrm{int}$ distribution in Section \ref{sec:discussion}. The limitations of the current work are discussed in Section \ref{sec:discussion2}, and our findings are summarized in Section \ref{sec:conclusion}.

\section{Lensing Basic}
\label{sec:theory}
\subsection{Lens equation and time-delay cosmography}
Gravitational lensing is the displacement of the image location which can be described by the lens equation:
\begin{equation}
\label{eq:lenseq_vector}
\beta = \theta - \alpha(\theta),
\end{equation}

where $\beta$ is source position and $\theta$ is image position in image plane, $\alpha(\theta)$ is the deflection angle at image position. $\alpha(\theta)$ is related to lensing potential via: 
\begin{equation}
\alpha(\theta)=\nabla \psi(\theta)
\end{equation}
and the relation between lensing potential and lensing convergence is:
\begin{equation}
\kappa(\theta)=\frac{1}{2} \nabla^2 \psi(\theta).
\end{equation}
where convergence is defined as:
\begin{equation}
\kappa(\theta) \equiv \frac{\Sigma(\theta)}{\Sigma_{\mathrm{cr}}}.
\end{equation}
The convergence is the surface mass density normalized by the critical lensing surface density
\begin{equation}
\Sigma_{\mathrm{cr}} \equiv \frac{c^2 D_{\mathrm{s}}}{4 \pi G D_{\mathrm{l}} D_{\mathrm{ls}}}
\end{equation}
In above equation, $c$ is the speed of light, $G$ is the gravitational constant, $D_{\mathrm{s}}$ is the angular diameter distance between the observer and the source, $D_{\mathrm{l}}$ is the angular diameter distance between the observer and the lens galaxy, and $D_{\mathrm{ls}}$ is the angular diameter distance between the lens galaxy and the source.

The time delay between two images A and B is given by:
\begin{equation}
\Delta t_{\Lambda \mathrm{B}}=\frac{D_{\Delta t}}{c}\left(\phi\left(\theta_{\mathrm{A}}, \beta\right)-\phi\left(\theta_{\mathrm{B}}, \beta\right)\right)
\end{equation}
where $\phi(\theta, \beta)$ is the Fermat potential \citep{1985A&A...143..413S,1986ApJ...310..568B}:
\begin{equation}
\phi(\theta, \beta)=\left[\frac{(\theta-\beta)^2}{2}-\psi(\theta)\right]
\end{equation}
and the time-delay distance $D_{\Delta t}$ \citep{1964MNRAS.128..307R,1992grle.book.....S,2010ApJ...711..201S} is:
\begin{equation}
D_{\Delta t} \equiv\left(1+z_{\mathrm{l}}\right) \frac{D_{\mathrm{l}} D_{\mathrm{s}}}{D_{\mathrm{Ls}}}
\end{equation}
Time-delay cosmography uses measured time delays with mass profiles from lens modeling and kinematics to constrain the time-delay distance, hence constraining the cosmological parameters. The Hubble constant is inversely proportional to the time-delay distance:
\begin{equation}
H_0 \propto D_{\Delta t}^{-1} .
\end{equation}

\subsection{Mass-sheet transform}
Constraining the mass profile of a lens galaxy is challenging due to the Mass-Sheet Transform (MST) \citep{1985ApJ...289L...1F}. The MST is defined by the equation:
\begin{equation}
\lambda \beta=\theta-\lambda \alpha(\theta)-(1-\lambda) \theta
\end{equation}
This transform is a multiplicative modification of the lens equation, preserving the image positions under a linear displacement of the source, $\beta \rightarrow \lambda \beta$. The term $(1 - \lambda)$ represents an infinite sheet of mass. Observables that are sensitive to the absolute source size, intrinsic magnification, or lensing potential can help break this degeneracy. 

The MST on convergence is:
\begin{equation}
\label{eq:mst}
    \kappa_{\lambda}(\theta) = \lambda \kappa(\theta) + 1 - \lambda
\end{equation}
where $\kappa_\lambda$ is the convergence after the mass-sheet transform. This equation indicates that if we scale the convergence by $\lambda$ and add a convergence sheet with $\kappa_=1-\lambda$, we would observe the same image as with the original convergence.

\begin{figure}
    \centering
    \includegraphics[width = 0.45\textwidth]{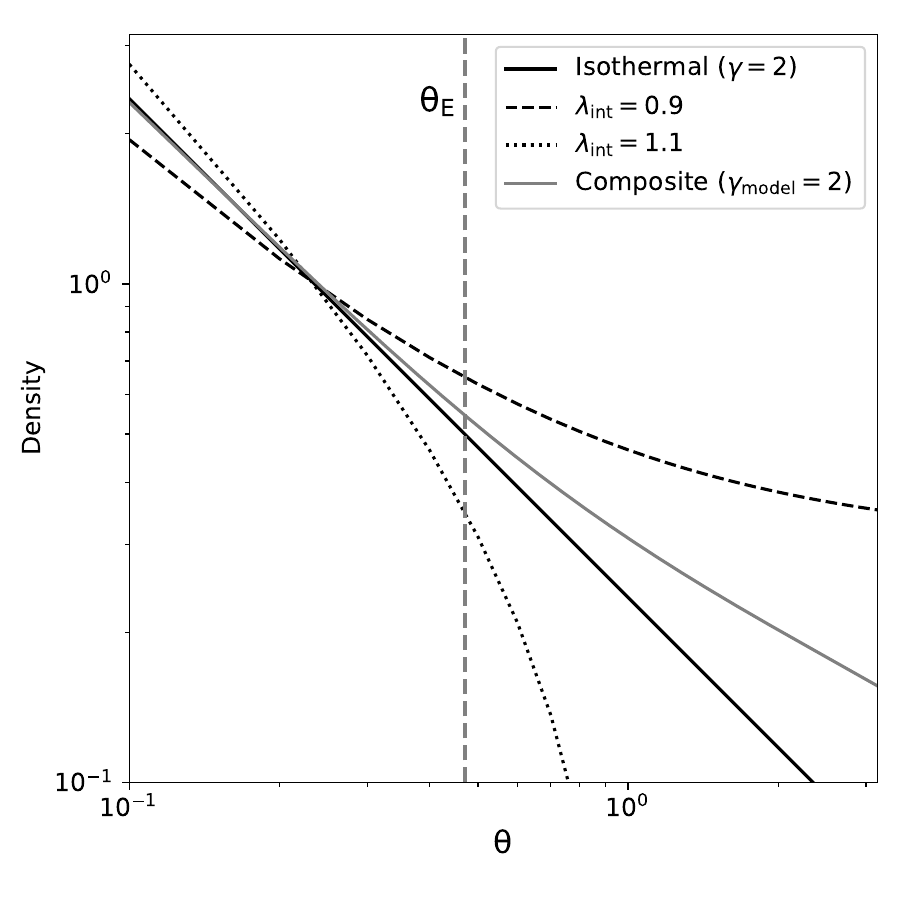}
    \caption{A illustration of different Mass-sheet transform on a power-law convergence profile. The y-axis is the density profile and the x-axis is the radius. The Einstein radius of this mock profile is 0.5$''$ and the power-law density slope $\gamma$ is 2. The gray profile is a star plus dark matter profile, with the same Einstein radius and effective density slope. The profile of a positive MST ($\lambda$<0) is more convex while the profile with a negative MST is more concave.}
    \label{fig:MST_illustration}
\end{figure}

Figure \ref{fig:MST_illustration} shows a power-law convergence profile with different MST. In general, a positive (1-$\lambda$) MST will make the profile more convex while a negative MST make it more concave. 

The mass-sheet transformed convergence will result in different time-delay:
\begin{equation}
D_{\Delta t \lambda}=\lambda^{-1} D_{\Delta t}
\end{equation}
hence, the Hubble constant also scale as: 
\begin{equation}
H_{0 \lambda}=\lambda H_0 .
\end{equation}
In this work, we use $\lambda_\mathrm{int}$ to represent the internal mass-sheet, which arises from deviations of the density profile from a power-law distribution.

\section{Lens Systems Simulation}
\label{sec:simulation}

\subsection{Deflector Population}
The deflector population is drawn from the CosmoDC2 catalog \citep{LSSTDarkEnergyScience:2019hkz}. The catalog is derived from the 'Outer Rim' simulation \citep{Heitmann:2019ytn}, a cosmological N-body simulation with a trillion particles within a 4.225 Gpc$^3$ volume. It spans 440 square degrees of the sky up to a redshift of \(z = 3\) and is complete to a magnitude depth of 28 in the r-band. We selected stellar mass and shape from the catalog. Instead of using the halo mass from the simulation, we tested two different stellar mass-to-halo mass relations. The deflector mass profile is a combination of a stellar component with a Sersic mass profile and a dark matter component with an NFW dark matter profile.

The Sersic profile \citep{1968adga.book.....S} discribes the surface mass density as:
\begin{equation}
\kappa(R) = \kappa_{\rm eff} \exp \left( -b_n \left[(\frac{R}{R_{\mathrm{eff}}})^{\frac{1}{n}}-1\right]\right)
\end{equation}
where n is Sersic index, $R_\mathrm{eff}$ is half-light radius, and $\kappa_\mathrm{eff}$ is the convergence at the half light radius. $b_n \approx 1.9992 * n - 0.3271$ is the approxminate solution of $\gamma\left(2 n ; b_n\right)=\frac{1}{2} \Gamma(2 n)$, where $\Gamma$ and $\gamma$ are respectively the Gamma function and lower incomplete Gamma function \citep{Ciotti:1999zs}. In this work, we assume that the mass of the baryon follows the light distribution. We can use this assumption for ellptical galaxies although it neglects the M/L gradient and the contribution from the gas and dust. 

The distribution of the dark matter is described by the Navarro, Frenk $\&$ White (NFW) profile \citep{Navarro:1995iw}:
\begin{equation}
\rho_{\mathrm{NFW}}(r) \equiv \frac{\rho_{\mathrm{s}}}{\left(r / r_{\mathrm{s}}\right)\left(1+r / r_{\mathrm{s}}\right)^2}
\end{equation}
where $\rho_{\mathrm{s}}$ is normalized density, and $r_s$ is the scale radius.

We randomly select early-type galaxies from the CosmoDC2 catalog with stellar masses larger than \(10^{11}M_{\odot}\) because most lens galaxies in the SLACS survey have stellar masses larger than that in \citet{Auger:2009hj}. Lower mass galaxies would have a small lensing cross-section and are less likely to host a source galaxy. We focus on early-type galaxies due to their high mass compared to other galaxy types. These galaxies are often targeted in lens-finding phases and constitute the main population of lens galaxies in current samples \citep{Bolton:2008xf, BOSS:2011bef, Gavazzi:2012kp, Sonnenfeld:2020xvy}.

The average Sersic index in the SLACS and BELLS samples is higher than that of typical early-type galaxies(around 5 and can be up to 10 \cite{BOSS:2012jco}). It is known that there are scaling relations between galaxy properties and the Sersic index \citep{Savorgnan:2013qsa}. In general, more massive galaxies tend to have a higher Sersic index, resulting in a steeper inner stellar density profile. We first obtained the linear relation of dark matter halo mass and velocity dispersion \citep{Zahid2018}:

\begin{equation}
\log (\frac{\sigma}{100 \mathrm{~km} \mathrm{~s}-1})= -0.007+0.3\log (\frac{M_{D M}}{10^{12} M_{\odot}}) + \mathcal{N}(0, 0.02)
\end{equation}
Then we fit a linear relation between the Sersic index and velocity dispersion from a sample of early-type galaxies in \cite{Hu:2008as}:
\begin{equation}
    n = 0.009\sigma+2.09 + \mathcal{N}(1, 0.06)
\end{equation}


To paint dark matter onto the stars, we used the Total Stellar-Halo Mass Relation (TSHMR) described in \citet{2020MNRAS.492.3685H} where the halo mass and virial mass scale as:
\begin{equation}
\label{eq:huangtshmr}
\log \mathcal{M}_{\star}^{\text {all }}=0.602_{-0.006}^{+0.005} \times\left(\log M_{\text {halo }}-13.5\right)+11.846_{-0.003}^{+0.003}
\end{equation}

Following \citet{Sonnenfeld2023} we then assume that the axis ratio of the dark matter halo is the same as the stellar component, and follows beta distribution $\mathrm{P}(q) \propto q^{5.28}(1-q)^{1.05}$. The concentration parameter of the NFW profile is obtained from \cite{2021MNRAS.506.4210I} using Colossus \citep{2018ApJS..239...35D}.

\subsection{Source Population}

Many methods have been used to discover gravitational lenses, and this has produced a heterogeneous sample with multiple discovery selection functions. The SLACS sample were detected as rogue emission lines at the wrong redshift in spectra of elliptical galaxies, whereas quasar lenses are typically discovered as bright multiply imaged quasars in imaging surveys. These different source populations form part of the difference in the selection function of the samples.


\subsubsection{[OII] emitter as galaxy-galaxy sources}

The typical type of source galaxy in SLACS survey is [OII] emitter due to the fact that the [OII] feature in the spectrum is very easy to identify. The redshift and luminosity distribution of [OII] emitter is adopted from the OII luminosity function used in \cite{Arneson2012}.  The [OII] luminosity function is fitted by a double power law:
\begin{equation}
\phi\left(L_{\mathrm{O}_{\mathrm{II}}}, z_s\right)=\frac{d N}{d L_{\mathrm{O}_{\mathrm{II}}}} \propto L_{\mathrm{O}_{\mathrm{II}}}^\alpha
\end{equation}
where \( L_{\mathrm{O}_{\text{II}}} \) is the \(\left[\mathrm{O}_{\mathrm{II}}\right]\) luminosity and \(\alpha\) is a dimensionless parameter. The low redshift (\(z < 0.6\)) luminosity function \(\left(\alpha_f, \alpha_b, \log L_{\mathrm{TO}}\right) = (-1.46, -2.90, 41.50)\) is obtained from \cite{Gilbank2010}, while the high redshift (\(z > 1\)) luminosity function \(\left(\alpha_f, \alpha_b, \log L_{\mathrm{TO}}\right) = (-1.27, -2.96, 41.91)\) is obtained from \cite{Zhu:2008ee}. We simulate [OII] emitters up to $z=1.5$. We also interpolate the turnover luminosity (\( L_{\mathrm{TO}} \)) in the power-law function as well as the slopes at brighter \(\left(\alpha_b\right)\) and fainter \(\left(\alpha_f\right)\) luminosities from \( L_{\mathrm{TO}} \) as a function of redshift. For each source galaxy, the axis-ratio and half-light radius are chosen from a random galaxy in CosmoDC2 with similar redshift. Then we draw 20 [OII] luminosity to increase the sample size.

\subsubsection{Galaxy-quasar sources}

We use the double power law for the quasar luminosity function from \cite{OM10} which has the following parametric form: 
\begin{equation}
\frac{\mathrm{d} \Phi_{\mathrm{QSO}}}{\mathrm{d} M}=\frac{\Phi_*}{10^{0.4(\alpha+1)\left(M-M_*\right)}+10^{0.4(\beta+1)\left(M-M_*\right)}}
\end{equation}
Where M refers to the absolute i-band magnitude of quasars. The faint-end slope is $\beta = -1.45$, while the bright-end slope is $\alpha = -3.31$ for redshifts lower than 3 and $\alpha = -2.58$ for redshifts greater than 3. The normalization of the luminosity function is fixed as $\Phi_*=5.34 \times 10^{-6} h^3 \mathrm{Mpc}^{-3}$. The break absolute magnitude $M_*$ is described as:

\begin{equation}
M_*=-20.90+5 \log h-2.5 \log f(z)
\end{equation}

where

\begin{equation}
f(z)=\frac{\mathrm{e}^{\zeta z}\left(1+\mathrm{e}^{\xi z_*}\right)}{\left(\sqrt{\mathrm{e}^{\xi z}}+\sqrt{\mathrm{e}^{\xi z_*}}\right)^2} \left(\zeta, \xi, z_*\right)=(2.98,4.05,1.60)
\end{equation}

To convert the i-band absolute magnitude to the apparent magnitude, we perform K-correction using a quasar template from \citet{2006ApJ...640..579G, 2001AJ....122..549V}. Then we assign a random galaxy in CosmoDC2 with similar redshift and use its ellipticity and scale radius. 

\subsection{Lensing Criteria}

\begin{figure*}
    \centering
    \includegraphics[width = 0.7\textwidth]{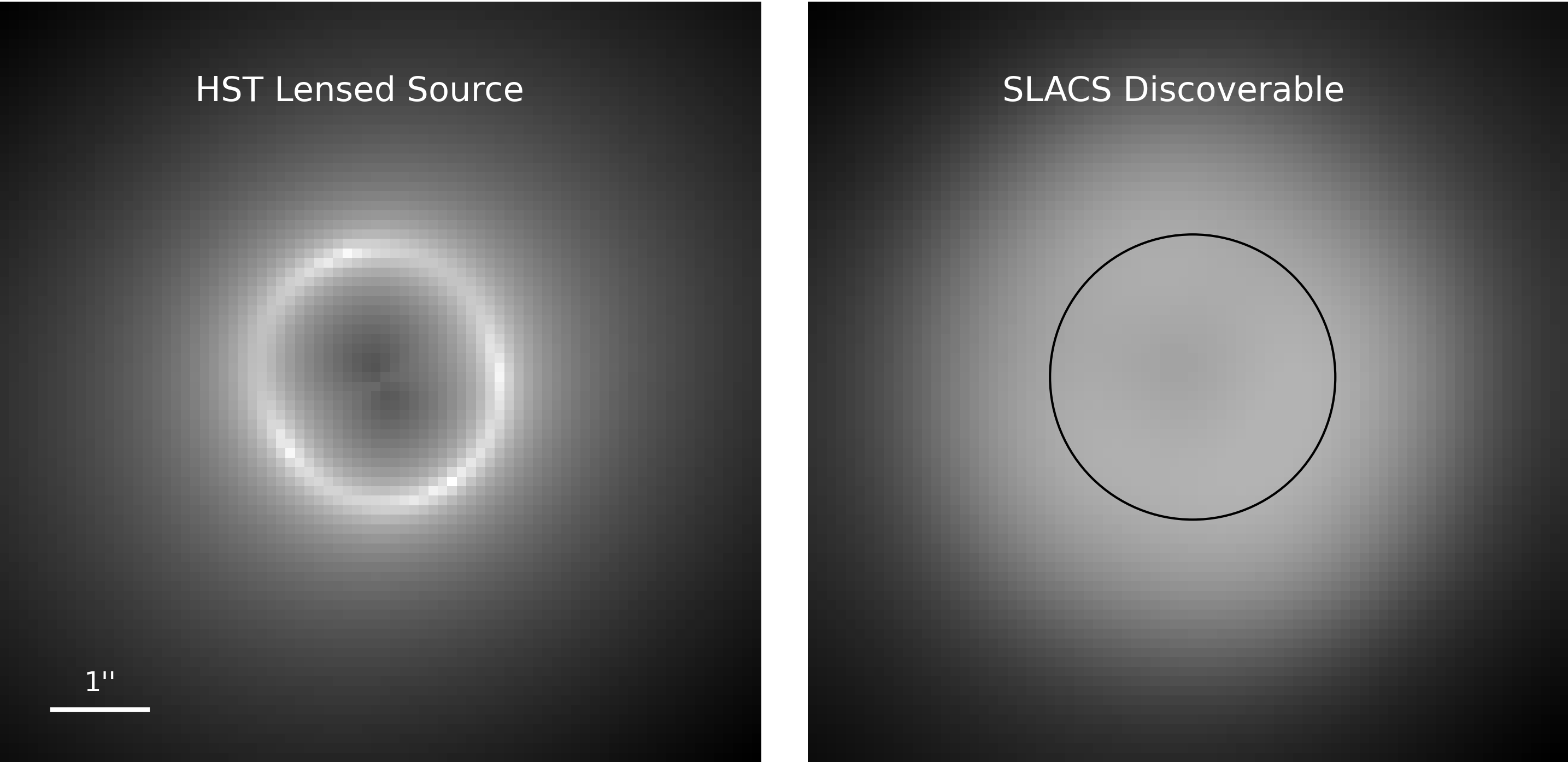}
    \includegraphics[width = 0.7\textwidth]{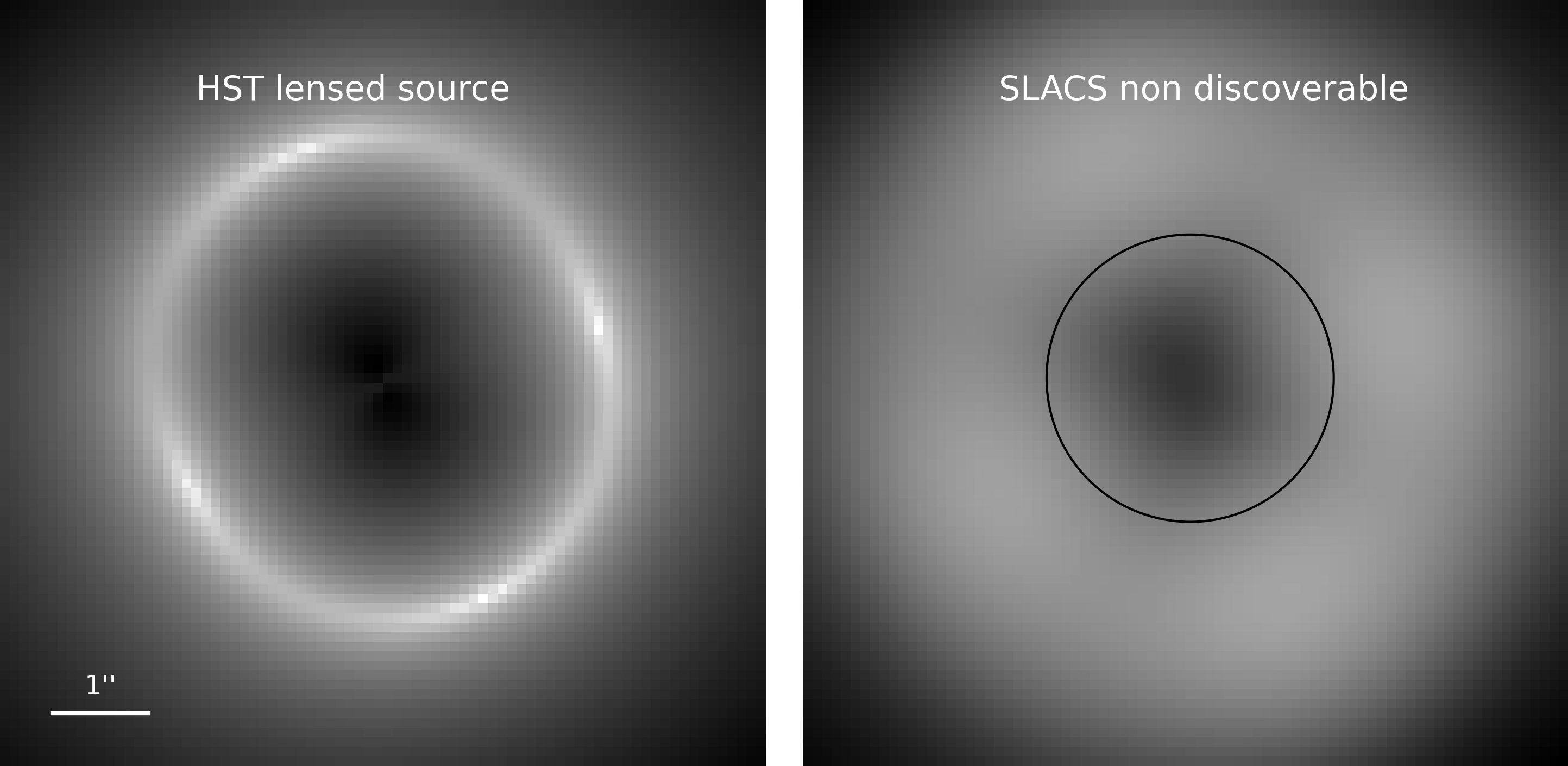}
    \caption{The example shows a SLACS lens (top figures) and a lens that will not be discovered in a spectroscopic survey (bottom figures). The left images are the lensed arc under HST resolution (FWHM $\approx$ 0.1$''$), and the right images are the lensed arc under SDSS resolution (FWHM $\approx$ 1.5$''$). The black circle represents an SDSS fiber with 3$''$ in diameter, only flux that falls within the fiber will be detected.}
    \label{fig:lens_condition}
\end{figure*}

\begin{figure*}
    \centering
    \includegraphics[width = 0.9\textwidth]{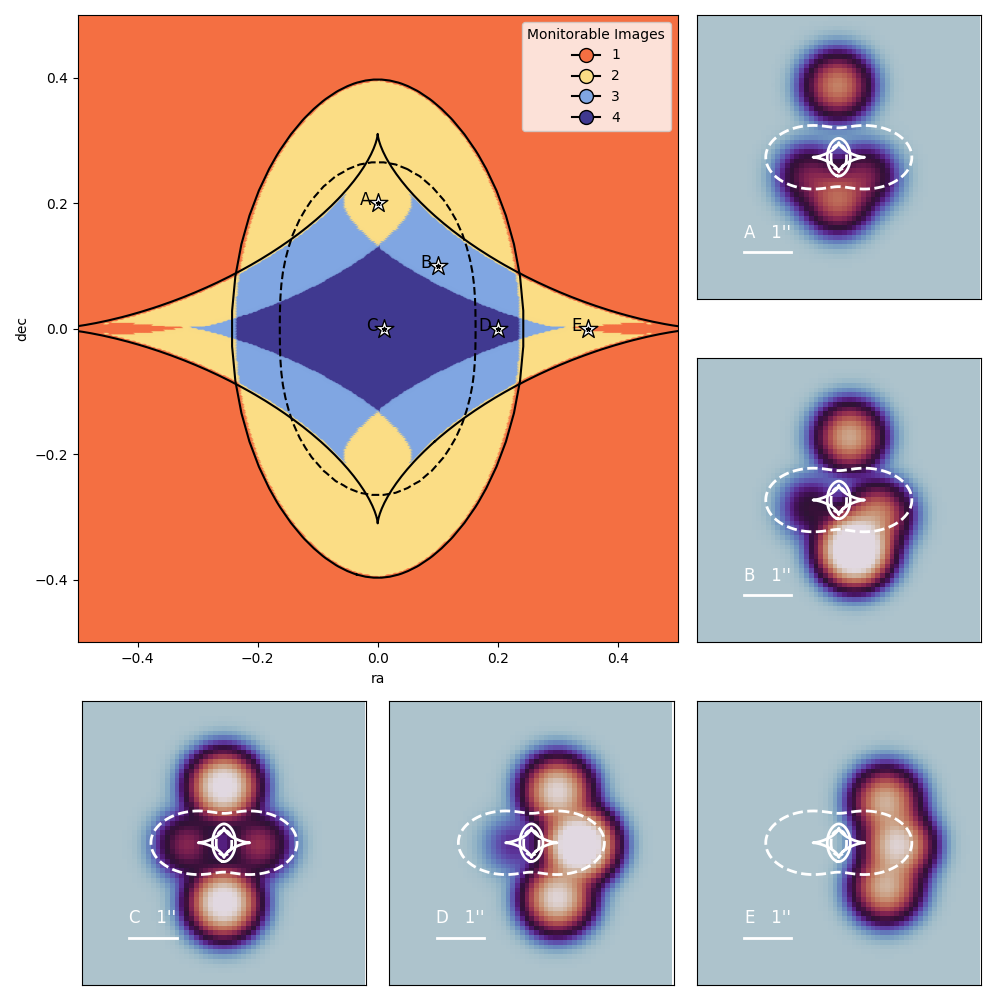}
    \caption{An example of the number of monitorable images given positions on the source plane is presented. The black solid line is the caustic and the dotted line is the critical curve. The example lens system has $\theta_E = 1''$, $\gamma = 1.6$, and an axis ratio of $q = 0.6$ ($\mathrm{e}_1 = 0.25$, $\mathrm{e}_2 = 0$). Note that in reality, this combination of $\gamma$ and $q$ is quite unlikely, but it provides a large region where the source is quadruply imaged, so we chose it for better illustration. The criterion for being a monitorable image here is that the distance between this image and all other images has to be 1$''$, and the image magnification has to be larger than 0.1. (Since the intrinsic luminosity of the quasar varies, For demonstration purposes this magnification cut only excludes images that are heavily demagnified. The actual selection is done by defining a magnitude cut) For images that are close to each other, they count as a single image. The surrounding figures are the corresponding quasar image configurations for source location A-E.}
    \label{fig:quasar_condition}
\end{figure*}

For each lens galaxy and source redshift, we calculate the Einstein radius of the system. The probability of a deflector having at least one source galaxy inside the multiply-imaged region is linearly proportional to the square of the Einstein radius. We use this weighting to draw lens and source pairs, with source galaxies randomly placed on the source plane such that the distance between the center of the lens and source galaxy is less than 1.5 times \(\theta_E\). Due to differences in the way they were discovered, we must have different selection criteria for time-delay lensed quasars, and spectroscopically discovered SLACS-type lenses. 

\subsubsection{Spectroscopically Selected Lenses}

For spectroscopically selected lenses (SLACS-type lenses), we convolve the image with a Gaussian PSF that has a FWHM of 1.5$''$ (the typical seeing of SDSS observations). Then, we calculate the [OII] flux within the circular aperture of 3$''$ diameter (the SDSS fiber). We follow the criteria of \cite{Dobler:2008qm}, where the [OII] flux should be larger than $6.0 \times$ $10^{-17}\,\mathrm{ergs} / \mathrm{s} / \mathrm{cm}^2$. We also require a minimum image magnification of three for the lens to exclude intrinsically bright [OII] emitters that would not be conclusively identified as lenses in the HST follow-up data. Figure \ref{fig:lens_condition} shows a lens system with a magnification of 7 (top panels) and 2.4 (bottom panels). The left figure shows the lensed source galaxy in HST resolution (FWHM $\sim$ 0.12$''$), and the right image shows the source galaxy in SDSS resolution (FWHM $\sim$ 1.5$''$). The black circle illustrates the SDSS fiber projected onto the sky. Following the SLACS data, we limit the lens galaxies to having a redshift smaller than 0.5 and the redshift of source galaxies are all below 1.2 \footnote{Our selection is specific to SLACS. The BELLS lenses \citep{Brownstein2012} will have a slightly different selection function since they were discovered in the BOSS survey which targeted higher redshift ellipticals and the spectrograph had a significantly smaller fiber size and higher CCD sensitivity.}.

In choosing targets for follow-up HST observation, the SLACS group added a further selection function. To maximize lensing rate, they targeted the systems with the largest expected Einstein radii, assuming an SIS profile and the measured SDSS velocity dispersion and redshifts. The smallest Einstein radius in \citetalias{Birrer2020} galaxy-galaxy lenses and galaxy-quasar lensed is 0.795$''$. Therefore, we drop all lenses with an Einstein radius smaller than 0.8$''$ from our sample \citep{2024arXiv240704771S}.

\subsubsection{Lensed Quasars}

For lensed quasars, we solve the lens equation and calculate the magnification of each quasar image. Figure \ref{fig:quasar_condition} is a toy model showing the number of images passing our criteria as a function of source plane position for an individual lens. We require that quad systems should have at least three images brighter than 21 magnitude in the $i$ band and separated by at least 1$''$'. Doubles should have both images satisfying that condition, we also treat quads with two images satisfying the criteria as doubles. This criterion is based on the properties of lensed quasar systems in \citetalias{Birrer2020}, monitored by Cosmograil \citep{Bonvin:2015jia}. Not all the sources within the diamond caustic can form a monitorable quad due to the magnification and image separation constraints.

\subsection{$\lambda_\mathrm{int}$ computation}

\begin{figure*}
    \centering
    \includegraphics[width = 0.8\textwidth]{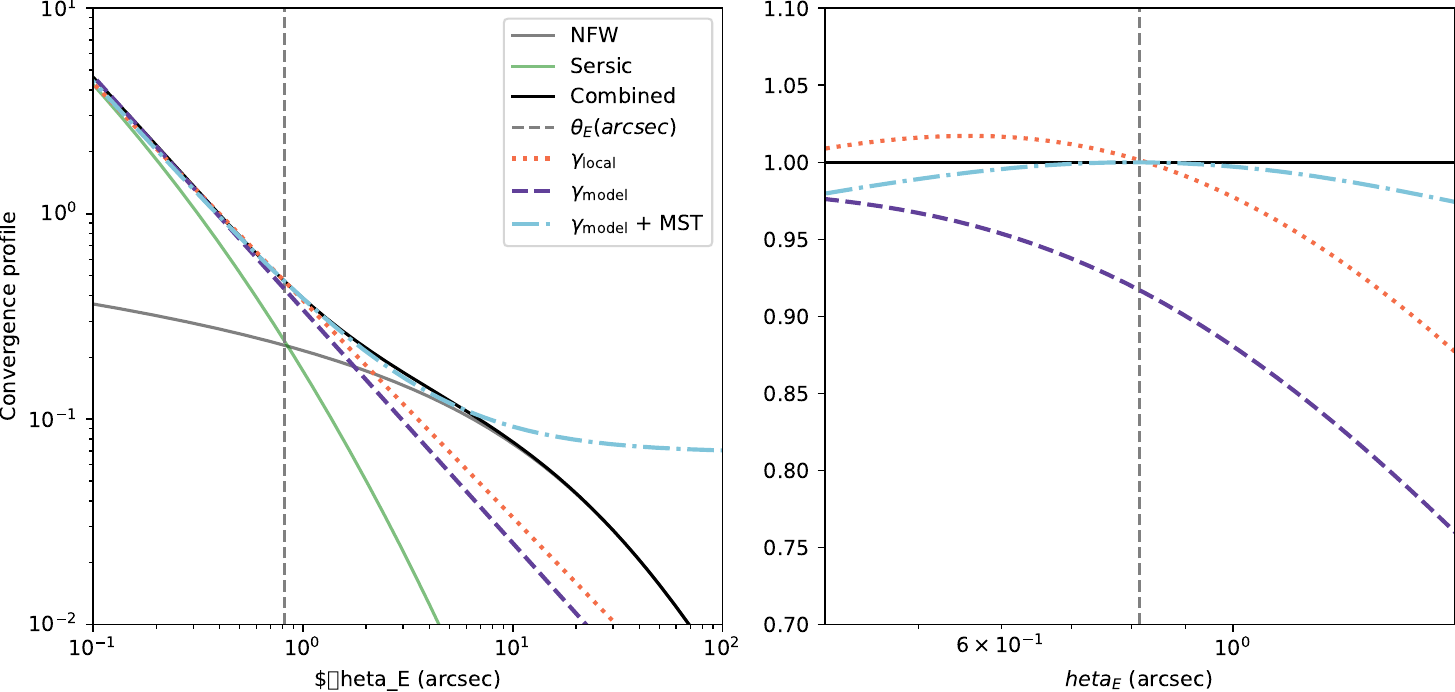}
    \caption{MST of a deflector galaxies' convergence profile. The lens redshift and source redshift are 0.2 and 0.8, respectively. The stellar component has $r_\mathrm{eff}$ = 8kpc, $n_\mathrm{sersic} = 0.6$, and $\mathrm{M}_\star = 10^{11.5} \mathrm{M}\odot$. The dark matter has a concentration of 5 and $\mathrm{M}_\mathrm{halo} = 10^{2} \mathrm{M}_\star$. The grey solid lines represent the dark matter NFW profile and the green profile is stellar Sersic profile. The black solid curve is the composite two-component profile. The orange dotted curve is the EPL profile marked as $\gamma_\mathrm{local}$ where the density slope is measured at the Einstein radius. The purple dashed line is the EPL profile where $\gamma_\mathrm{model}$ is measured by computing the MST invariant quantity. The cyan dashed line is the modeled profile with a mass-sheet transform using $\lambda$ from equation \ref{eq:mst}. The left and right panels show the actual convergence profile, and the normalized convergence profile (each divided by the composite profile).}
    \label{fig:mst_transfer_example}
\end{figure*}

To simulate the $\lambda_\mathrm{int}$ measurement in \citetalias{Birrer2020}, we first need to measure the logarithmic density slope $\gamma$ that would be recovered from lens modeling. The Elliptical power-law (EPL) profile is a typical profile that describes the lensing mass distribution \citep{1994A&A...284..285K, Barkana:1998qu}. The convergence $\kappa$ at a given angular position $\mathrm{\theta}$ is expressed by the equation:
\begin{equation}
\label{eq:epl}
\kappa\left(\theta_1, \theta_2\right)=\frac{3-\gamma}{2}\left[\frac{\theta_{\mathrm{E}}}{\sqrt{q \theta_1^2+\theta_2^2 / q}}\right]^{\gamma-1}
\end{equation}
where \( \gamma \) is the logarithmic density slope of the profile, q is the axis ratio of the minor and major axes, and \( \theta_E \) is the Einstein radius. Previous research has demonstrated that lens modeling does not measure the actual density slope at the Einstein radius \citep{Schneider:2013sxa,Sonnenfeld:2017dca, Kochanek:2019ruu, Kochanek:2020crs, Birrer:2021iuz}. Instead, the measured quantity is the radial stretch of the arc, which can be expressed by an MST invariant quantity: 
\begin{equation}
\frac{\partial_r \lambda_{\mathrm{rad}}\left(\theta_{\mathrm{E}}\right)}{\lambda_{\mathrm{rad}}\left(\theta_{\mathrm{E}}\right)},
\end{equation}
where $\lambda_{\mathrm{rad}}^{-1}$ is the radial component of the eigenvalues of lensing Jacobian matrix. $\lambda_{\mathrm{rad}}$ corresponds to the stretch factor of the
source in radial direction and can also be expressed as the derivative of the image position $\theta_r$ with respect to the source position $\beta_r$ in radial direction: 
\begin{equation}
\lambda_{\mathrm{rad}}=\frac{\partial \theta_\mathrm{rad}}{\partial \beta_\mathrm{rad}}.
\end{equation}
For elliptical power-law model, the MST invariant quantity at Einstein radius is:
\begin{equation}
\label{eq:mstinvariant}
\frac{\partial_r \lambda_{\mathrm{rad}}\left(\theta_{\mathrm{E}}\right)}{\lambda_{\mathrm{rad}}\left(\theta_{\mathrm{E}}\right)}=\frac{\gamma_\mathrm{model}-2}{\theta_{\mathrm{E}}}
\end{equation}
where $\gamma$ is the logarithmic density slope from lens modeling. Detailed derivation can be found in \cite{Birrer:2021iuz}, and equivalently in \cite{Kochanek:2019ruu,Sonnenfeld:2017dca}. Rearranging equation \ref{eq:mst}, we have:
\begin{equation}
\lambda_\mathrm{int} = \frac{\kappa_{\lambda} - 1}{\kappa - 1}
\end{equation}

Figure \ref{fig:mst_transfer_example} shows the 2D radial convergence profiles of a mock lens galaxy and the different density profiles that would be inferred by fitting a powerlaw density profile to the data.  The profile from fitting a pure powerlaw deviates substantially from the truth, whereas EPL plus a MST transform is a much closer approximation of the true profile around the Einstein radius.

\section{Results}
\label{sec:result}
\begin{figure*}
    \centering
    \includegraphics[width = \textwidth]{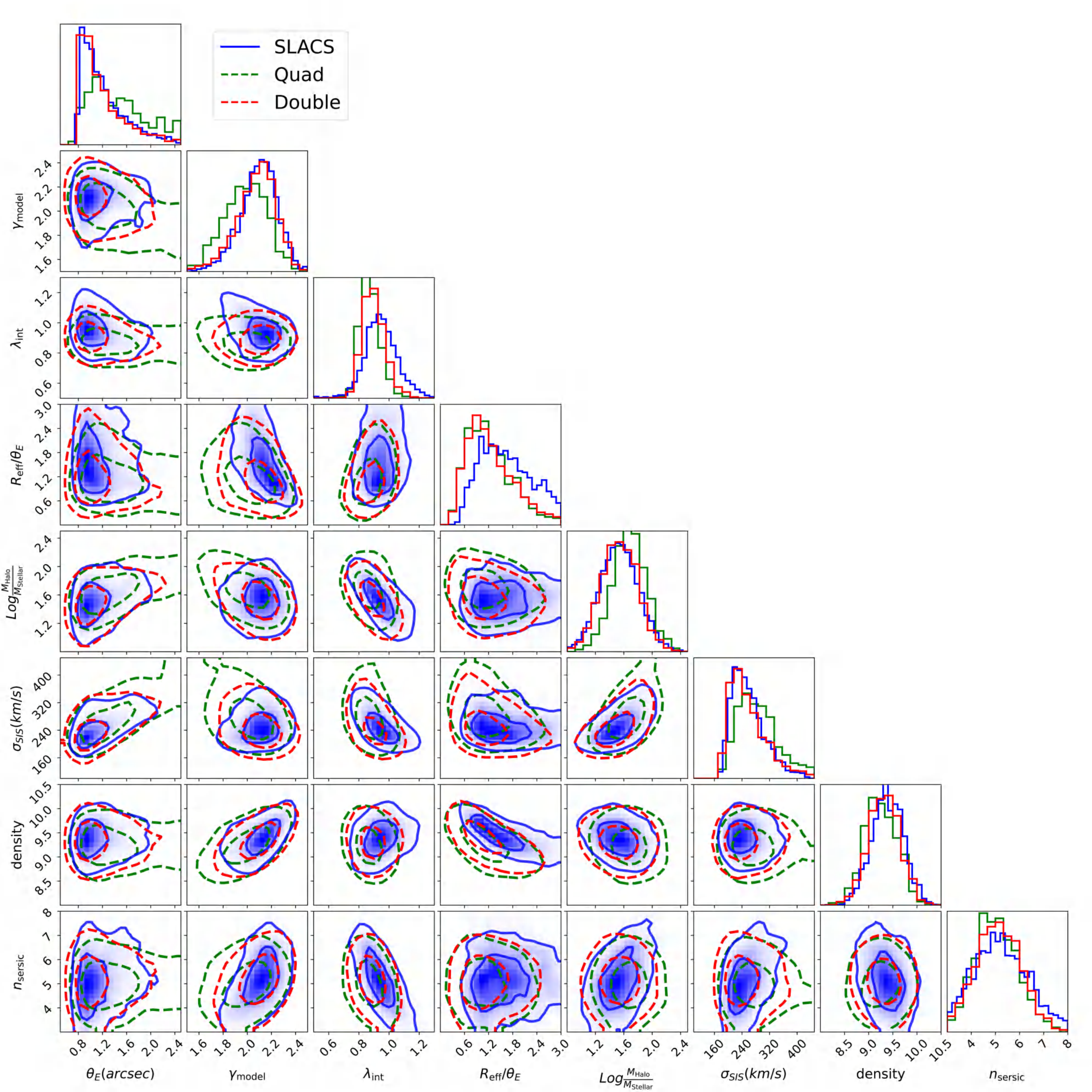}
    \begin{tabular}{ccccccccccc} 
    \hline
    Name & $\theta_E ('')$ & $\gamma_{\text{local}}$ & $\gamma_\mathrm{model}$ & $\lambda_\mathrm{int}$ & $R_\mathrm{eff}(kpc)$ & $M_\star (M_\odot)$ & $M_{DM}(M_\odot)$ & $log\frac{M_\mathrm{DM}}{M_\star}$ & $\sigma_{SIS} (km/s)$ & $n_{\text{sersic}}$\\
    \hline
    Double & $0.75^{+0.50}_{-0.19}$ & $2.02^{+0.13}_{-0.16}$ & $2.12^{+0.13}_{-0.15}$ & $0.92^{+0.07}_{-0.07}$ & $6.3^{+4.4}_{-2.4}$ & $11.4^{+0.3}_{-0.2}$ & $12.8^{+0.5}_{-0.4}$ & $1.40^{+0.28}_{-0.25}$ & $215^{+62}_{-34}$ & $5.04^{+0.71}_{-0.69}$\\
    Quad & $1.53^{+0.99}_{-0.52}$ & $1.82^{+0.19}_{-0.16}$ & $1.96^{+0.19}_{-0.20}$ & $0.87^{+0.07}_{-0.06}$ & $10.9^{+7.0}_{-4.5}$ & $11.8^{+0.3}_{-0.3}$ & $13.6^{+0.5}_{-0.5}$ & $1.72^{+0.26}_{-0.27}$ & $298^{+129}_{-62}$ & $5.08^{+0.75}_{-0.72}$\\
    SLACs & $1.11^{+0.47}_{-0.22}$ & $2.05^{+0.13}_{-0.14}$ & $2.11^{+0.13}_{-0.14}$ & $0.95^{+0.09}_{-0.08}$ & $6.8^{+5.2}_{-2.9}$ & $11.6^{+0.3}_{-0.3}$ & $13.1^{+0.5}_{-0.5}$ & $1.5^{+0.26}_{-0.25}$ & $248^{+48}_{-35}$ & $5.04^{+0.79}_{-0.77}$\\
    \hline
    \end{tabular}
    \caption{The distribution of SLACS lenses (blue), monitorable quads (green), and monitorable double (red). SLACS samples are selected such that the Einstein radius has to be larger than 0.8$''$ and must be modelable. The bottom table shows the corresponding distribution.}
    \label{fig:corner}
\end{figure*}

\begin{figure*}
    \centering
    \includegraphics[width = \textwidth]{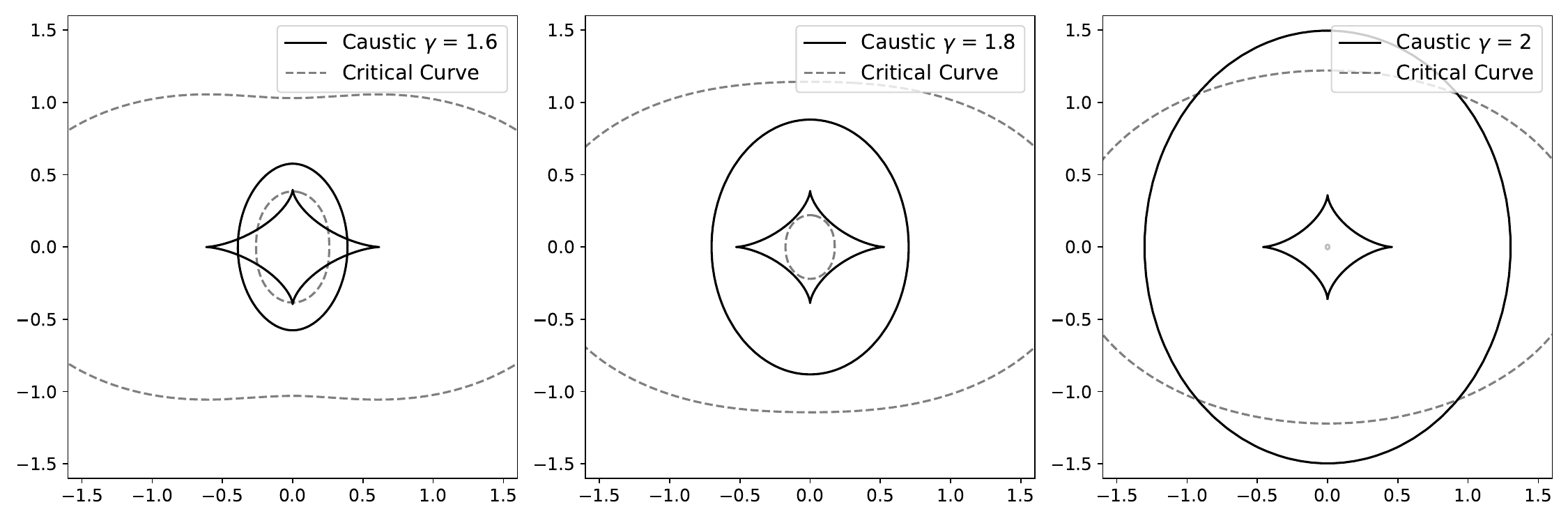}
    \caption{The caustics and critical curves of a power-law elliptical lens. The values of the parameters are set to $\theta_E$ = 1.5, q = 0.66 (e1 = 0.2, e2 = 0.0). It shows that increasing $\gamma$ mildly decreases the diamond caustic but significantly increases the elliptical caustic. Indicating a steeper density profile would produce more doubles and fewer quads. }
    \label{fig:caustic}
\end{figure*}

In this section, we present the resulting lens population of galaxy-quasar and galaxy-galaxy lenses. Figure \ref{fig:corner} shows the relevant lens parameters in our sample, the distribution of each parameter is listed in the table below. 
The most significant feature of quads is that they have shallower density profiles ($\gamma_\mathrm{model}$), the relative shift between monitorable quads and input ensemble in our sample is smaller than the result from \citet{Collett:2016muz}, this is due to the fact that \citet{Collett:2016muz} assumed that SLACS lenses were the parent population of deflectors for lensed quasars, whereas in this work we have constructed a single parent population of deflectors and investigated both selection functions. Essentially, \citet{Collett:2016muz} measures the selection function for lensed quasars, whereas we are primarily interested in the difference in the selection functions of SLACS and lensed quasars.


Figure \ref{fig:caustic} shows the caustic and critical curve of a deflector galaxy under different power-law index with an Einstein radius of 1.5$''$ and axis ratio of 0.66. The surface area of the central diamond region (where the source will be quadruply imaged) slightly decreases with increasing power-law slope. In contrast, the outer elliptical caustic (where the source will be doubly imaged) significantly increases with the power-law index and remains its size after $\gamma = 2$. So instead of saying quads prefer shallower profiles, one may conclude that doubles prefer steeper profiles. 

Due to the image separation requirement, the monitorable quads systems have a larger Einstein radius on the population level compared to doubles and SLACS. As a result, quads also have a larger stellar mass and halo mass. The ratio between dark matter halo mass and stellar mass (log $\frac{\mathrm{M}_\mathrm{halo}}{\mathrm{M}_\star}$) is larger than the other two populations. This is because a higher log $\frac{\mathrm{M}_\mathrm{halo}}{\mathrm{M}_\star}$ produces a shallower profile, and quads have shallower profiles. Another reason for quads having larger halo mass is that the Einstein radius has a weak positive correlation to log$\frac{\mathrm{M}_\mathrm{halo}}{\mathrm{M}_\star}$. This relation is partly due to the TSHMR we used where a system with a larger halo mass usually has a larger log $\frac{\mathrm{M}_\mathrm{halo}}{\mathrm{M}_\star}$. Another reason is that since the halo mass is much larger than the stellar mass, the Einstein radius is more sensitive to the fractional change in halo mass (except for lenses at very low redshift where the stellar mass dominates the mass inside the Einstein radius). 

For all populations, we also measured the $R_\mathrm{eff}/\theta_E$ of each system. Because $\lambda_\mathrm{int}$ is a local quantity that should vary with the position of the Einstein ring within the lens galaxy. The SLACS population has a higher $R_\mathrm{eff}/\theta_E$. The limiting selection criteria of SLACS lenses is the [OII] flux within the 3$''$ SDSS fibre aperture. This limits the Einstein radius of the lens, which combined with the SLACS bias towards low redshift lenses drives to larger $R_\mathrm{eff}/\theta_E$. The variation of $\lambda_\mathrm{int}$ with $R_\mathrm{eff}/\theta_\mathrm{E}$ is also accounted for in TDCOSMO IV, allowing a degree of freedom for $\lambda_\mathrm{int}$ in this direction.

\subsection{The mass sheet parameter}

Based on our assumed galaxy models that combine stellar mass (represented by a Sersic profile) and dark matter halos (modeled with an NFW profile), we find in our simulations that the $\lambda_\mathrm{int}$ has mean values less than 1 (approximately 0.9) across all populations. This result is consistent with the findings of \citetalias{Birrer2020}. As illustrated in Figure \ref{fig:mst_transfer_example}, the Sersic + NFW composite profile generally produces a convex mass distribution curve. Consequently, when fitting the mass distribution with a power-law profile, the modeled convergence value at the Einstein radius is about 5$\%$ higher than the actual convergence from our simulated galaxies, leading to $\lambda_\mathrm{int} < 1$. However, as demonstrated in section \ref{sec:gnfw}, increasing the inner slope of the dark matter profile will raise the $\lambda_\mathrm{int}$

The parameter $\lambda_\mathrm{int}$ is covariant with the Sersic index and $\log \left(\frac{\mathrm{M}\mathrm{halo}}{\mathrm{M}_\star}\right)$, as these two parameters directly determine the shape of the profile. Higher Sersic indices and higher values of $\log \left(\frac{\mathrm{M}\mathrm{halo}}{\mathrm{M}_\star}\right)$ result in a lower $\lambda_\mathrm{int}$.  Given that monitorable quads favor a shallower profile, they also tend to prefer higher halo masses, thereby leading to a lower $\lambda_\mathrm{int}$ ($0.87^{+0.07}_{0.06}$ compared to $0.95^{+0.09}_{-0.08}$) with similar scatter.  

\begin{figure}
    \centering
    \includegraphics[width=0.5\textwidth]{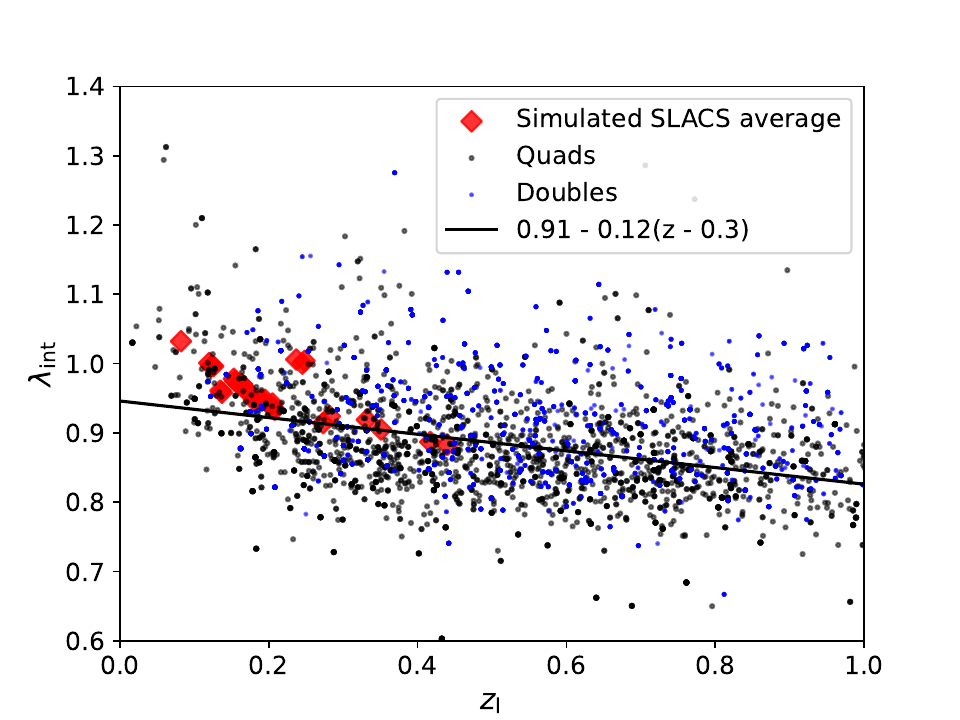}
    \caption{$\lambda_\mathrm{int}$ as a function of lens redshift for simulated SLACS lenses and monitorable quads. The black line is the linear fit to the quads data points.}
    \label{fig:enter-lambda_vs_zl}
\end{figure}

\subsection{Debiasing $H_0$ from TDCOSMO and SLACS}
\label{sec:tdcosmo}

To understand the impact of the different $\lambda_\mathrm{int}$ of our simulated lensed quasars and SLACS lenses, we repeat the selection process of \citetalias{Birrer2020}.

The selection of \citetalias{Birrer2020} is: 
\begin{itemize}
\item The deflector galaxies need to be elliptical galaxies.  
\item The lenses should have good-quality models from an automated and uniform modeling procedure. 
\item The Singular Isothermal Sphere (SIS) equivalent velocity dispersions $\sigma_\mathrm{sis}$ need to lie within the range of [200, 350] km/s. 
\end{itemize}

Since our simulated lenses are all ellipticals and we have perfect knowledge of the mass profiles, we need only add the $\sigma_\mathrm{SIS}$ cut. We additionally require the lensed quasar redshifts to be below 2.5, as in the \citetalias{Birrer2020} sample.

\cite{Knabeletal2024} studied  14 SLACS lenses with Keck Cosmic Web Imager (KCWI) on the Keck Telescope. They found that the most recent velocity dispersion estimates from SDSS fiber spectra are biased by a factor of $\sim$5.3$\%$ with respect to KCWI data integrated within the same aperture. \citetalias{Birrer2020} used the systematically lower SDSS velocity dispersions. The new KCWI velocity dispersions are consistent with SLACS IX \citep{slacs9} measured velocity dispersions, and SLACS IX has velocity dispersion measurements for all 33 of the SLACS sample used in \citetalias{Birrer2020}. We therefore use the velocity dispersions of \citet{slacs9}.

The lensed quasar population in \citetalias{Birrer2020} consists of 6 quads and one double. We randomly pair up the lensed quasars and SLACS lenses in our sample, matching each of the real lenses with a simulated twin with the same $R_\mathrm{eff}/\theta_E$. This allows us to realize the difference in $\lambda_\mathrm{int}$ between mock TDCOSMO and SLACS lenses as used in \citetalias{Birrer2020}. We ran the pairing process 1000 times to reduce the shot noise. To see how this propagates into $H_0$ we  rerun the TDCOSMO notebook\footnote{\url{https://github.com/TDCOSMO/hierarchy_analysis_2020_public/}} from \citetalias{Birrer2020}, but rescaling the $\lambda_\mathrm{int}$

The resulting $H_0$ is $63.6^{+4.8}_{-3.9}$ km s$^{-1}$Mpc$^{-1}$ if we only account for the selection effect, compared to the $64.6^{+4.4}_{-4.5}$ km s$^{-1}$Mpc$^{-1}$ for the same dataset in \citetalias{Birrer2020} but accounting for the velocity dispersion bias. 

\label{sec:gnfw}

\begin{figure}
    \centering
    \includegraphics[width = 0.5\textwidth]{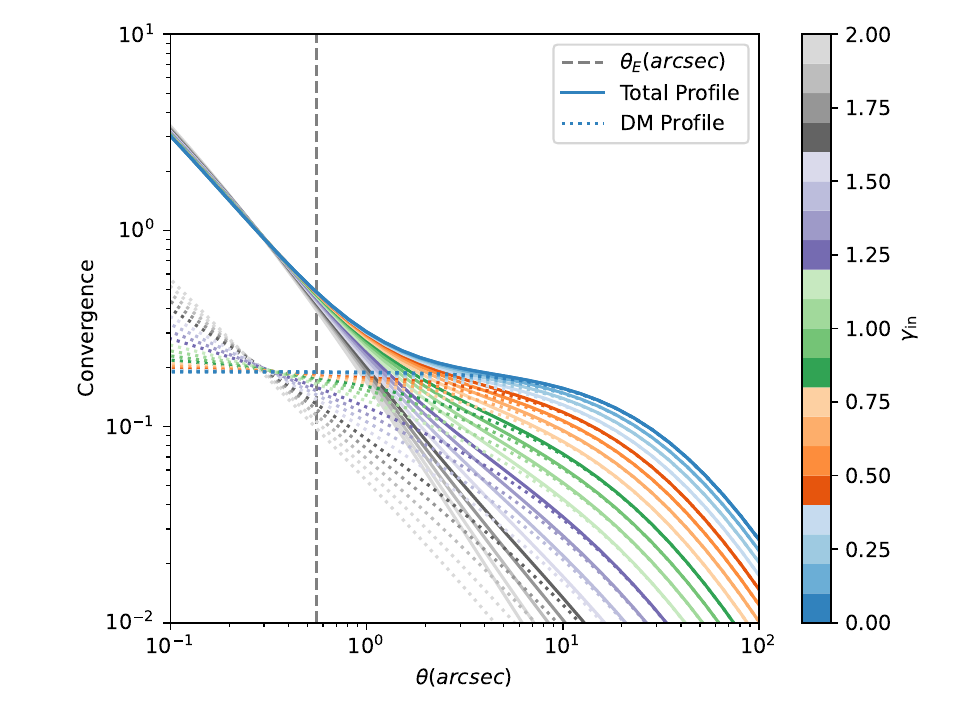}
    \includegraphics[width = 0.45\textwidth]{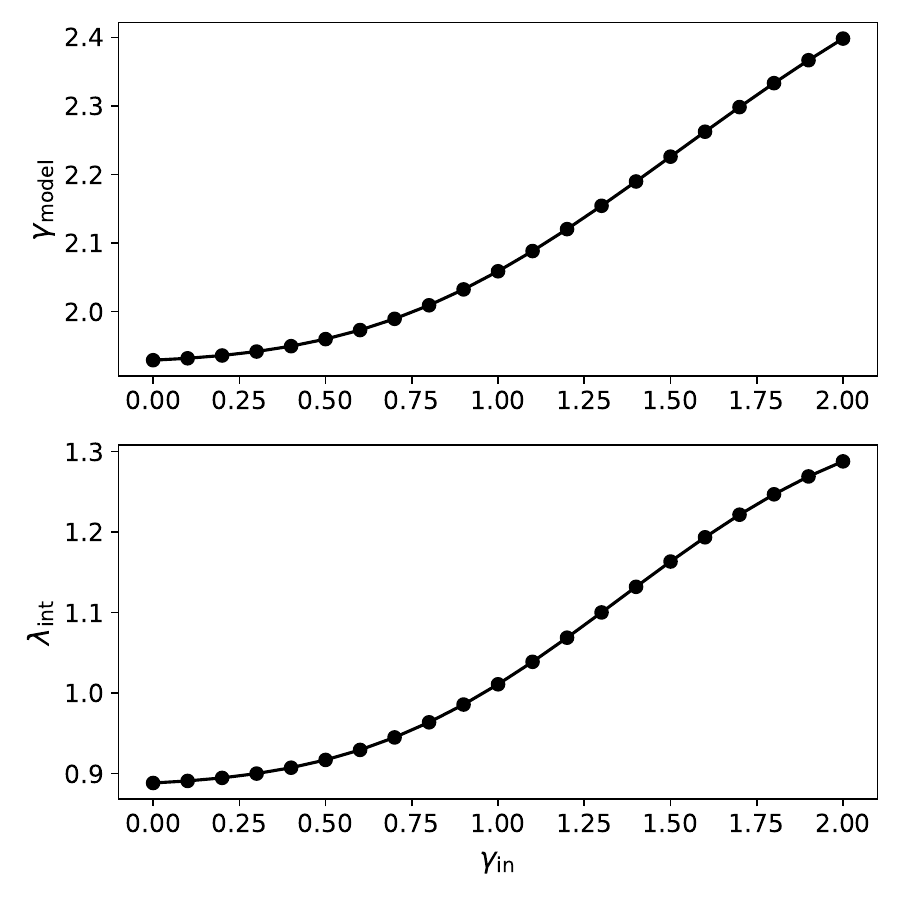}
    \caption{Top panel: The convergence profile of a simulated lens system with fixed stellar component and different dark matter inner-slope $\gamma_\mathrm{in}$ (from 0 to 2). The convergence at the scale radius is adjusted so that for each $\gamma_\mathrm{in}$, the Einstein radius remains unchanged. The dashed line is the Einstein radius where the $\lambda_\mathrm{int}$ is measured. Middle panel: $\gamma_\mathrm{model}$, and $\lambda_\mathrm{int}$ as a function of $\gamma_\mathrm{in}$.}
    \label{fig:innerslopevsgamma-label}
\end{figure}



\section{Limitations}
\label{sec:discussion}

\begin{figure}
    \centering
    \includegraphics[width=0.5\textwidth]{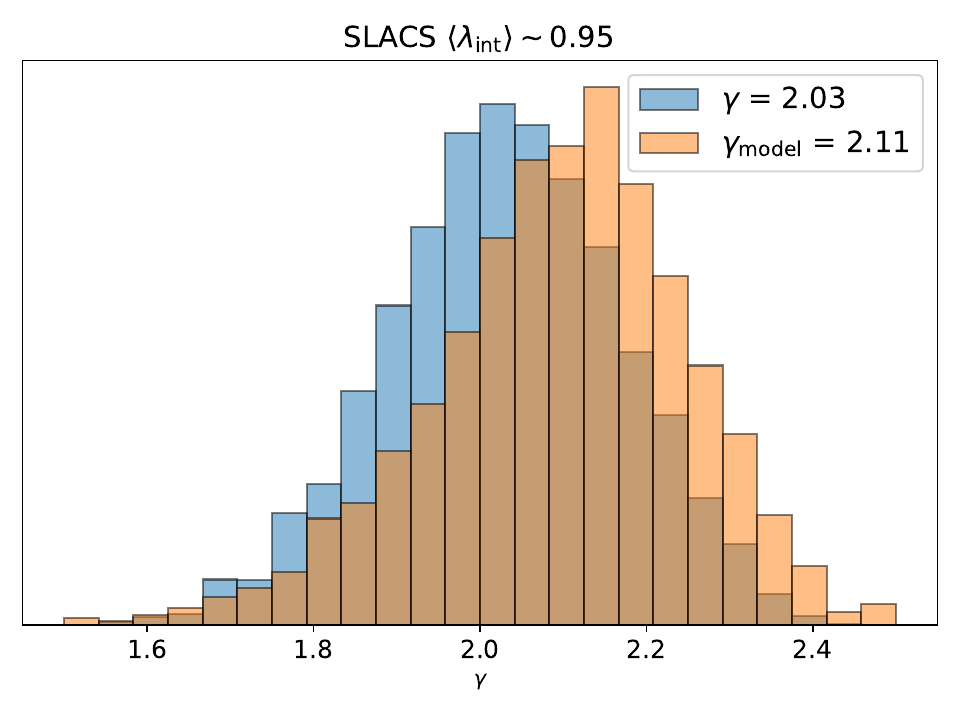}
    \caption{The distribution of the density slope at the Einstein radius ($\gamma$) and modelled density slope ($\gamma_\mathrm{model}$) of SLACS lenses.}
    \label{fig:density_slope}
\end{figure}

\begin{figure*}
    \centering
    \includegraphics[width=\textwidth]{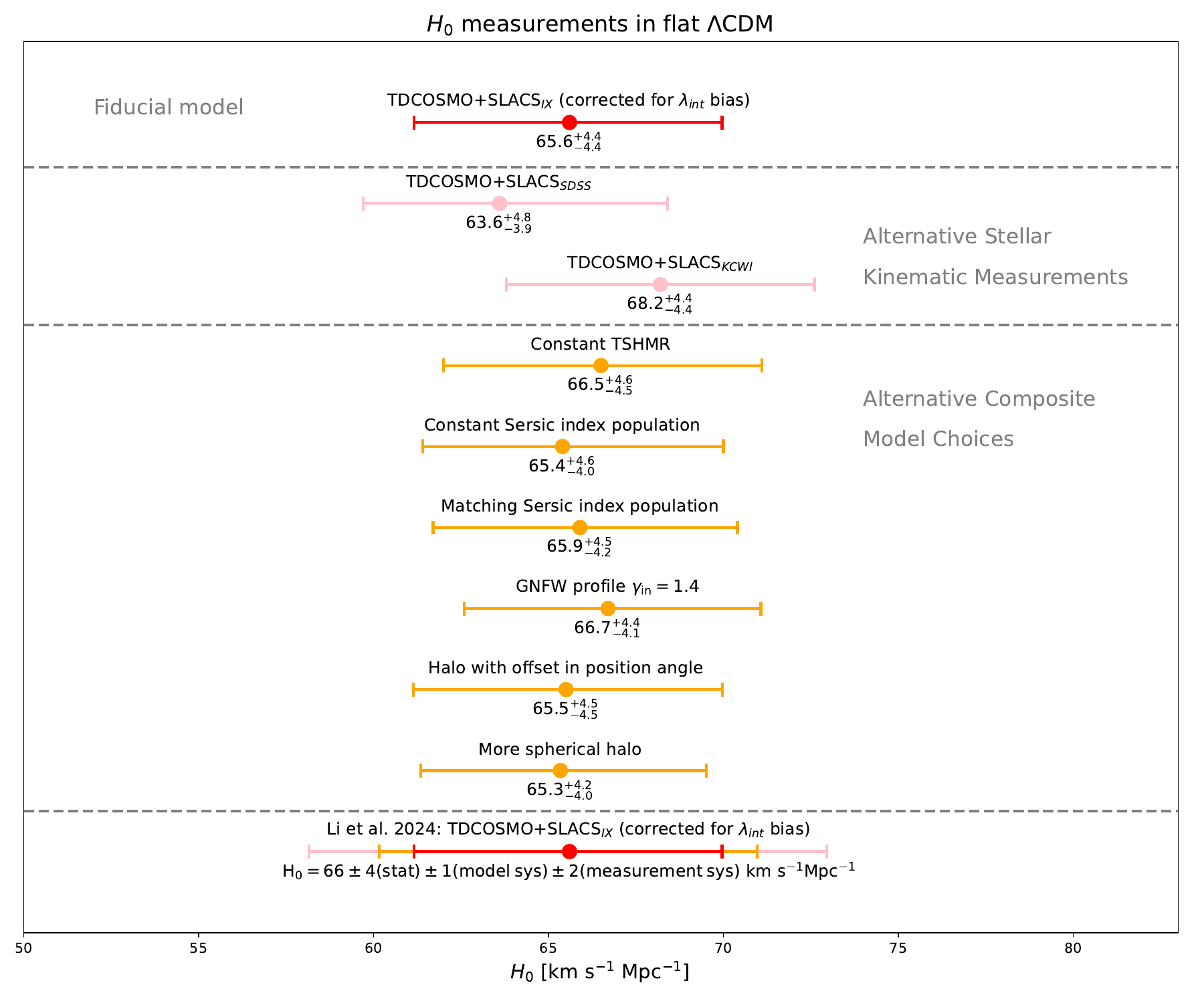}
    \caption{Impact of various systematic choices on the inferred $H_0$ from the combined SLACS and TDCOSMO dataset. We generate mock SLACS-like galaxy-galaxy lenses and TDCOSMO-like galaxy-quasar lenses and simulate the selection effect between their mass-sheet parameter $\lambda_\mathrm{int}$. Then we run \citetalias{Birrer2020} hierarchical inference code taking the selection effect into account to obtain the new $H_0$ measurement. The measurement shown in red at the top row shows our fiducial result, where we adopt our baseline composite deflector population and velocity dispersion measurements (see Sections \ref{sec:simulation} and \ref{sec:tdcosmo}). The \textbf{pink} measurements illustrate the effect of using alternative velocity dispersion measurements (Comparing SLACS IX values with those from SDSS or KCWI), which shifts the $H_0$ estimate by a few km\,s$^{-1}$\,Mpc$^{-1}$. The Orange measurements display the impact of different choices in constructing the composite deflector population (such as variations in the stellar-to-halo mass relation or galaxy model parameterizations). The Bottom row is our final $H_0$ result obtained after combining these systematic variations. This summary panel shows that although each individual systematic choice produces a modest shift in $H_0$, the overall effect remains within our quoted uncertainties.}
    \label{fig:h01}
\end{figure*}

\begin{figure*}
    \centering
    \includegraphics[width=\textwidth]{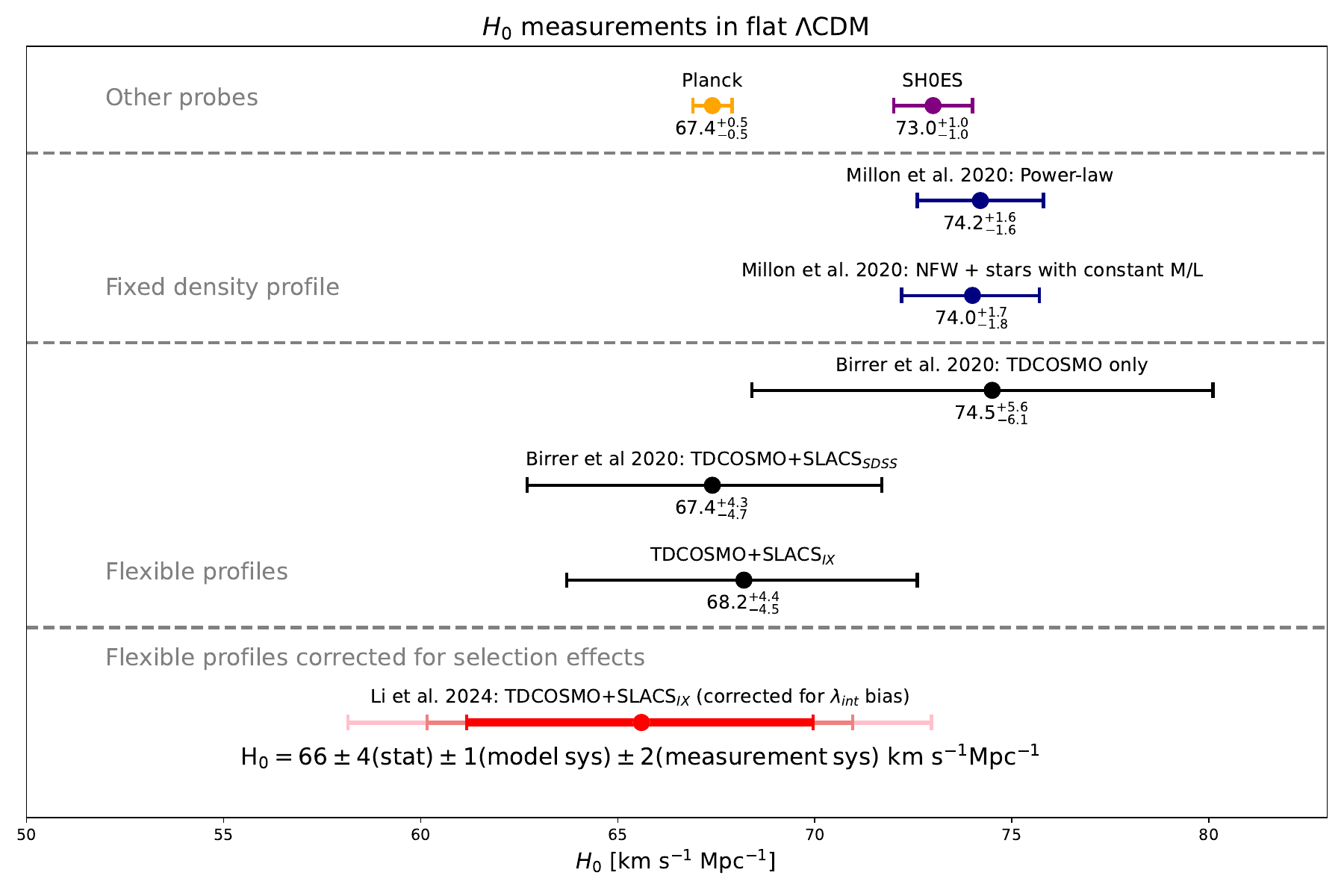}
    \caption{$H_0$ measurement from different methods. The top row is from the CMB \citep{2020A&A...641A...6P} and the cosmic distance ladder \citep{Riess2022}, whilst the other results are for different approaches with lensed quasars. The blue results assume we know the true density profiles of lenses, black allows a flexible density profile but assumes SLACS lenses and lensed quasars are intrinsically the same. Red shows our work accounting for the difference in selection function between SLACS and TDCOSMO lenses. Pink shows the systematics of our result associated with how we built our composite model and how velocity dispersions are measured.}
    \label{fig:h02}
\end{figure*}

In this work, we have built a population of deflectors, modelled the selection function of galaxy-galaxy and galaxy-quasar lenses and seen how this maps into different values of the Hubble constant when run through the methodology of \citetalias{Birrer2020}.

In our lens population $\lambda_\mathrm{int}$ decreases with lens redshift and increases with $R_{\mathrm{eff}} / \theta_{\mathrm{E}}$. Analysing the real SLACS population, \cite{Tan2024} also found that the $\lambda_\mathrm{int}$ depends on the redshift of the lens galaxy as $\mu_{\lambda_{\text {int }}}=$ $0.91_{-0.09}^{+0.10}-0.12_{-0.26}^{+0.31}(z-0.3)-0.04_{-0.04}^{+0.04}\left(R_{\mathrm{eff}} / \theta_{\mathrm{E}}-1\right)$. At redshift 0.3 and $R_\mathrm{eff}/\theta_E = 1$, the $\lambda_\mathrm{int}$ of both populations in our sample has a mean value of 0.91 which is in agreement with these two research. Similar to \cite{Tan2024}, We find that $\lambda_\mathrm{int}$ decreases with redshift. Figure \ref{fig:enter-lambda_vs_zl} shows $\lambda_\mathrm{int}$ as a function of lens galaxy redshift. This is the main reason that although our monitorable doubles share similar $\gamma_\mathrm{model}$ distribution with SLACS lenses, its $\lambda_\mathrm{int}$ distribution is lower. 

While our $\lambda_\mathrm{int}$ prediction for the mock SLACS population is consistent with \citetalias{Birrer2020} and \cite{Tan2024} at z $\sim$ 0.3 and $r_\mathrm{eff}/\theta_E = 1$, this work is still limited by various assumptions we made. In this section, we investigate how choices in our simulations can impact the infered value of the Hubble constant. The results of making different plausible choices are shown in Figure \ref{fig:h01}.

\subsection{Stellar mass profile}

The single Sersic profile without mass-to-light gradient we used in this work is a simplified version of the baryon mass distribution. It is not a perfect fit to observed lens light profiles \citep{2013ApJ...766...70S}. Modeling lens light with double Sersic or multi-gaussian \citep{He:2024udi} profiles usually results in better residuals. Therefore, we do not anticipate being able to establish a direct comparison between the intrinsic $\lambda_\mathrm{int}$ measurements from our mock galaxy systems and those of the real lens systems presented in \citet{Birrer2020}. The choice of a single Sersic profile is simply because we have better knowledge of its population distribution and its correlation with other lens parameters. This parameterization can also describe the concaveness of the stellar distribution which is the key factor that determined the $\lambda_\mathrm{int}$. As shown in previous sections, the shape of the stellar profile also strongly correlates with $\lambda_\mathrm{int}$, so any unknown evolution related to stellar distribution will result in an evolution in $\lambda_\mathrm{int}$ as well. For example, \cite{BOSS:2012jco} found that the Sersic index marginally increases with redshift. Yet many studies reveal that the Sersic index of elliptical galaxies decreases with redshift (e.g. \citep{2013MNRAS.428.1460B, 2024MNRAS.527.6110O}). 

To test the impact of the sersic index, we generate a lens sample assuming the sersic index population is invariant of dark matter halo mass. After running the \citetalias{Birrer2020} analysis, the $H_0$ of this sample is: $65.4^{+4.6}_{-4.0}$ km s$^{-1}$Mpc$^{-1}$.

Since the Sersic index correlates with the $\lambda_\mathrm{int}$,  we fit a single Sersic profile to the TDCOSMO lensed quasar with a mask on the arc. The fitting is performed in F814W band using {\sc{pysersic}} \citep{pysersic}. The resulting distribution is $n_\mathrm{sersic} \sim 4\pm 2$. We draw lens samples so that the distribution of their Sersic index agrees with TDCOSMO IV. With this sample, the resulting $H_0$ is $65.9^{+4.5}_{-4.2}$ km s$^{-1}$Mpc$^{-1}$.

The mass profile in our sample tends to be shallower for a larger Einstein radius. This is because galaxies with more mass have larger halo masses, and the Einstein radius is more sensitive to fractional changes in halo mass. A larger $\frac{M_\mathrm{halo}}{M_\star}$ results in a shallower density profile. This is also shown in \cite{newman2015}; as the Einstein radius increases, the deflector properties exhibit behavior more similar to galaxy clusters. However, the currently existing galaxy-scale lens sample does not show this behavior. This is because the current lens sample is too small. Other effects include that the shallower mass profile leads to a smaller total caustic area, as shown in figure \ref{fig:caustic}, so a larger halo mass would help a system produce a larger lensing cross-section, the resulting shallower mass profile counteracts this effect and makes the cross-section smaller. Some other systematics might also affect our results: most of the current strong lensing deflector density slopes are measured from lens modeling (except for SL2S lenses). And $\gamma_\mathrm{model}$ does not scale with $\theta_E$. The halo mass of early-type galaxies at the high mass end might not be well understood, yet this relation between Einstein radius and density profile mostly depends on the relation between halo mass and stellar mass. 

\subsection{Stellar to Halo Mass relation}
\label{sec:tshmr}

The distribution of mass parameters for galaxy-quasar versus galaxy-galaxy samples with constant TSHMR (same as figure \ref{fig:corner}) is shown in figure \ref{app:constantMh}. The parameters show various behaviors compared to samples with varying TSHMR. But the $\lambda_\mathrm{int}$ distribution of two samples remains less than 1. Which is in good agreement with \citetalias{Birrer2020} and \cite{Tan2024}.

Compare to observations: the population distribution of $\gamma_\mathrm{model}$ for SLACS lenses in both of our samples is $2.11^{+0.13}_{-0.14}$ and $2.12^{+0.14}_{-0.13}$, which is in good agreement with the measured value of approximately 2.08 $\pm$ 0.13 \citep{2021shajib}. The $\gamma_\mathrm{model}$ for quads from the varying TSHMR is $1.96^{+0.19}_{-0.20}$, while for the fixed TSHMR it is $2.05^{+0.14}_{-0.13}$. The quad population from the fixed TSHMR is closer to the distribution of 30 quads from the pixel level modelling in \cite{Stride}, which is 2.05 ± 0.09. This does not imply a preference for one TSHMR over the other, because our $\gamma_\mathrm{model}$ comes from equation 26 and the modeling pipeline in \cite{Stride} used a tight and informative prior on $\gamma_\mathrm{EPL}$. Instead, it demonstrates that our measurement of $\gamma_\mathrm{model}$ is realistic at the population level, thereby validating the computation of $\lambda_\mathrm{int}$.

We run the \citetalias{Birrer2020} with constant TSHMR samples and get a H0 of: $66.5^{+4.6}_{-4.5}$ km s$^{-1}$Mpc$^{-1}$. This slight increase in the Hubble constant is because in this SLACS and TDCOSMO populations have become more similar in the mean $\lambda_\mathrm{int}$. This is because SLACS and TDCOSMO populations have different $\frac{M_\mathrm{halo}}{M_\star}$ ratio in our fiducial TSHMR model, while in a constant TSHMR sample, the $\frac{M_\mathrm{halo}}{M_\star}$ ratio between two samples is more similar, resulting in more similar $\lambda_\mathrm{int}$ at the population level.

\subsection{The dark matter density profile}

\cite{2021MNRAS.503.2380S} showed that SLACS lenses are well described by the NFW profile. Multiple previous studies have suggested that the inner slope of the NFW profile of massive elliptical galaxies has a wide range of values\citep{1998ApJ...499L...5M, 1999MNRAS.310.1147M, 2000ApJ...529L..69J, 2001ApJ...549L..25K, 2001ApJ...555..504W, 2018MNRAS.476..133O}. The Generalized NFW profile is parameterized as:
\begin{equation}
\rho=\frac{\rho_s}{\left(r / r_s\right)^\gamma_\mathrm{in}\left(1+r / r_s\right)^{3-\gamma_\mathrm{in}}},
\end{equation}
where $\gamma_\mathrm{in}$ is the inner slope. The distribution of the inner slope is not well understood, \cite{Sheu2024} hierarchy modeled 58 slacs lenses and obtained a mean value of $\gamma_\mathrm{in}$ $\sim$ 1.20, \cite{Oldham2018} modeled 12 massive early-type galaxies with a mean value of $\gamma_\mathrm{in} \sim 1.80$, and the simulation from \cite{Sonnenfeld2023} shows that the $\gamma_\mathrm{in}\sim 1.40$ for their assumed population. To compare with \cite{Sonnenfeld2023}, we set the distribution to $\gamma_\mathrm{in}$ = $1.4 \pm 0.2$ and assume there is no correlation with other parameters. Then we approximate the convergence profile with multi-gaussian ellipse \citep{2019MNRAS.488.1387S}, and redo the above analysis. We found that the $\gamma_\mathrm{in}$ distribution for lensed quasar slightly favors a lower value (1.36 $\pm$ 0.2) compared to SLACS lenses (1.39 $\pm$ 0.2). Figure \ref{fig:innerslopevsgamma-label} shows the convergence profile with different inner slopes and the covariance between $\gamma_\mathrm{in}$ and $\gamma_\mathrm{model}$. Lower $\gamma_\mathrm{in}$ produces a shallower density profile, but this correlation is not very strong since the dark matter is sub-dominant compared to stellar components within the Einstein radius. Combining all of these effects nets to a small increase in the Hubble constant: The $H_0$ result from galaxy population with GNFW profile is $66.7^{+4.4}_{-4.1}$ km s$^{-1}$Mpc$^{-1}$.

One of the uncertainties in the mass profile lies in the response of the dark matter inner profile to the halo mass and baryon component. However, the inner slope distribution is not well understood, and the simulations have lots of uncertainty. Whether the difference of $\lambda_\mathrm{int}$ is true can be verified by observations. As \cite{2024arXiv240401513K} shows, the average MSD of 500 galaxies can be constrained with weak lensing observation to $\sim$ 10$\%$ with HSC level data, and 1-3$\%$ with LSST level data.

Our fiducial model assumes that the dark matter halo and the stellar component share the same axis ratio and position angle. Although this assumption was also used in \cite{Gomer2022} and \cite{Sonnenfeld2023}, it is not clear whether it introduces additional systematic biases. Previous studies have shown that the position angle and ellipticity of the light and dark components of galaxies do not match \citep{Bruderer2016,Sheu2024}. We investigated a model that incorporates an offset in the position angle between the dark matter and stars, as well as a model in which the dark matter is more spherical than the stars. Following the results of \cite{Bruderer2016}, the offset angle is assumed to follow a normal distribution centered at 0 with a standard deviation of 0.6 rad. In the more spherical model, the ellipticity of the dark matter is set to 30\% of that of the stars. We find that both modifications have a negligible impact on the inferred value of $H_0$ ($65.5^{+4.5}_{-4.5}$ km s$^{-1}$ Mpc$^{-1}$ and $65.4^{+4.2}_{-4.0}$ km s$^{-1}$ Mpc$^{-1}$, compared to the fiducial $H_0$ of $65.6^{+4.4}_{-4.4}$ km s$^{-1}$ Mpc$^{-1}$).

\subsection{Correlations between $\lambda_\mathrm{int}$ and other parameters}

In this study, we determine the $\gamma_\mathrm{model}$ by evaluating the radial derivative of the deflection angle at the Einstein radius to compute the modeled power-law density slope. Mathematically, the invariant in equation \ref{eq:mstinvariant} is correct (e.g. \cite{Kochanek:2020crs}), but in reality, the result from the equation \ref{eq:mstinvariant} does not always agree with the result of lens model \citep{Etherington2023}. Equation \ref{eq:mstinvariant} only represents a first-order approximation in lens modeling because the actual lens modeling constrains derivatives of the lens potential at the location of lensed images, which sometimes does not coincide with the Einstein radius. One may choose to weight the lensed image by its pixel value and compute the average derivative of the deflection angle on a lens-by-lens case, which can be more accurate but computationally expensive. Furthermore, variations in source reconstruction strategies and incorrect priors on the source or lens model parameterization can influence the $\gamma_\mathrm{model}$. In this paper, We are making the strong assumption that the modelling in TDCOSMO is not systematically biased. As demonstrated by \cite{2024arXiv240608484G}, for different lensing algorithms on a mock system with sufficient SNR, the recovered $\gamma_\mathrm{model}$ consistently exhibits bias in a similar direction across different methods. Therefore, we believe that our $\gamma_\mathrm{model}$ for both mock lensed quasars and mock slacs lenses aligns well with most major lens modeling algorithms.

The primary reason for the variation in $\lambda_\mathrm{int}$ observed in this study is the difference in $\gamma_\mathrm{model}$, which is influenced by changes in the $\frac{M_\mathrm{halo}}{M_\star}$ ratio. \cite{Xu:2015dra} also found that a lower slope results in a lower $\lambda_\mathrm{int}$, as a larger $\frac{M_\mathrm{halo}}{M_\star}$ makes the convergence profile more concave. We measured the density slope at the Einstein radius in our mock lens by fitting a power-law slope around the Einstein radius and compared it with the modeled density slope. Figure \ref{fig:density_slope} shows ths $\gamma$ and $\gamma_\mathrm{model}$ distribution of SLACS lenses. A $\lambda_\mathrm{int}$ of 0.95 results in the $\gamma$ shift of $4\%$. 

In addition, the location of the lensed image with respect to the lens galaxy ($r_\mathrm{eff}/\theta_E$) also leads to the variation in $\lambda_\mathrm{int}$. This variation comes from the shape of the convergence profile having different concaveness at different $r_\mathrm{eff}/\theta_E$. And lens modelling is only sensitive to the profile around the position of lensed images. Since simulations struggle to reproduce the actual mass profile of the galaxies, and the exact profile is also hard to measure by observations. It is hard to study the correlations between $r_\mathrm{eff}/\theta_E$ and $\lambda_\mathrm{int}$ analytically. \cite{Tan2024} found that the linear dependency of $\lambda_\mathrm{int}$ on $r_{\mathrm{eff}} / \theta_{\mathrm{E}}$ is close to 0, with a measured value of $-0.04_{-0.04}^{+0.04}$. Our galaxies with a GNFW profile show similar results, where the linear dependency is approximately $0.02$.


\section{Looking to the future}
\label{sec:discussion2}
In this section, we briefly look forward to how the mass sheet degeneracy can be expected to impact future analyses combining galaxy-galaxy and galaxy-quasar lenses. In the future most galaxy-galaxy lenses will be discovered in imaging rather than spectroscopic surveys \citep{Collett:2015roa}, and lensed quasar populations will be intrinsically fainter \citep{OM10}.

\subsection{QSO magnitude cut}
\begin{figure}
    \centering
    \includegraphics[width=0.5\textwidth]{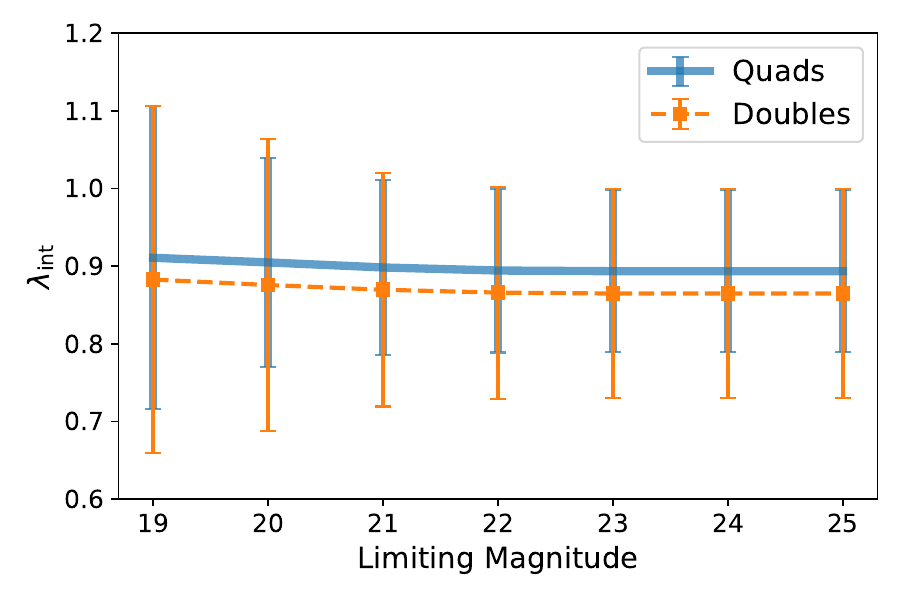}
    \caption{The mean $\lambda_\mathrm{int}$ varying as a function of the faintest quasar image magnitude}
    \label{fig:magnitudecut}
\end{figure}

LSST is able to monitor quasars with fainter luminosities and new deconvolution algorithms like STARRED allow us to relax the image separation requirement \citep{Michalewicz2023, 2024AJ....168...55M}. Figure \ref{fig:magnitudecut} shows the mean $\lambda_\mathrm{int}$ varying as a function of the faintest quasar image magnitude. For both Quads and Doubles, $\lambda_\mathrm{int}$ slowly decreases. This is because when we relax the image magnitude requirement, we are actually probing quasars at higher redshift. This drives lens galaxy to higher redshift as well, according to figure \ref{fig:enter-lambda_vs_zl}, this will slighly decrease the $\lambda_\mathrm{int}$ on the population level.

\subsection{Galaxy-Galaxy lenses discovered in LSST and Euclid}
\label{sec:Galaxy-Galaxy lenses discovered in LSST and Euclid}
\begin{figure}
    \centering
    \includegraphics[width=0.45\textwidth]{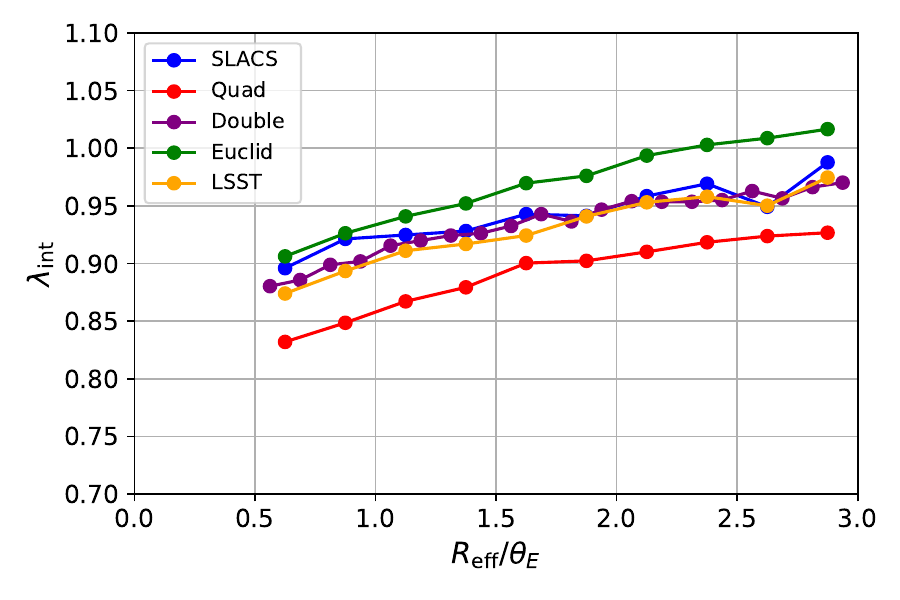}
    \caption{$\langle\lambda_\mathrm{int}\rangle$ as a function of $R_\mathrm{eff}/\theta_E$ for different types of lens populations.}
    \label{fig:eff}
\end{figure}

To understand the selection function of galaxy-galaxy lenses discovered in imaging surveys like Euclid and LSST, we generate a population of lensed extended sources. The source galaxy is randomly selected from CosmoDC2 \citep{LSSTDarkEnergyScience:2019hkz} with the criterion that its redshift must be greater than that of the deflector galaxy. 

We use the lensing Criteria from \cite{Collett:2015roa}, which requires that the magnification of the source galaxy to be at least three, this means that the surface area of the lens is magnified by the factor of 3, a lensed arc would be visible under this criteria. Another requirement is that the magnitude of the lensed source galaxy should be brighter than the detection threshold. We also require that the image and counter-image must be resolved. This can be expressed as: 
\begin{equation}
\theta_E^2>r_s^2+s^2 / 2
\end{equation}
where $s$ is the size of PSF (for ground based survey, this is close to the seeing) and $r_s,\left(x_s, y_s\right)$ is the unlensed source size. Then we require that the tangential shearing of the arcs must be detectable, and the source should be detected with sufficient signal-to-noise for a human or lens-finding algorithm to recognize as a lens. This means that the magnification should satisfy:
\begin{equation}
\mu_{\mathrm{TOT}} r_s>s,
\end{equation}
and the total signal to noise should be larger than 20:
\begin{equation}
\mathrm{SNR}_{\mathrm{TOT}}>20.
\end{equation}

Figure \ref{fig:eff} shows the $\lambda_\mathrm{int}$ vs $R_\mathrm{eff}/\theta_E$ for different population and model. The LSST population has similar $\lambda_\mathrm{int}$ as SLACS and doubles, Quads has a lower $\lambda_\mathrm{int}$ and Euclid population has a higher $\lambda_\mathrm{int}$. The reason Euclid has higher $\lambda_\mathrm{int}$ is that the Euclid population has lower mass deflector, so they have smaller $\frac{M_\mathrm{halo}}{M_\star}$ according to our fiducial TSHMR, and hence lower $\lambda_\mathrm{int}$.

\section{Conclusions}
\label{sec:conclusion}

In this work, we generate mock SLACS galaxy-galaxy lenses and TDCOSMO lensed quasars. We use a stars-plus-dark-matter mass profile for the parent population of deflectors in the Universe and then model the selection function of SLACS and TDCOSMO-like populations. We compute the $\lambda_\mathrm{int}$ of each mock lens when modeled with a power law mass profile. We found that a power law plus MST profile is a good approximation to our underlying two-component mass profile. Our model reproduces the observed trends in the density profile slope of SLACS lenses that has been seen in \cite{Tan2024}, giving confidence that our underlying deflector population and SLACS selection function is a realistic approximation of the real Universe.

At fixed $R_\mathrm{eff}/\theta_E$, quadruple imaged quasars have a  $\lambda_\mathrm{int}$ which is approximately 0.06 lower than a SLACS lens or a double imaged quasar system. However, SLACS lenses typically have higher $R_\mathrm{eff}/\theta_E$ than quasars and so the average $\lambda_\mathrm{int}$ for both populations is approximately 0.9. This supports the $H_0$ shift from the H0LiCOW value to the lower value in \citetalias{Birrer2020}. However, based on the latest KCWI study, the SLACS velocity dispersion used in \citetalias{Birrer2020} is biased by 5.3$\%$, correcting this bias alone will give us $H_0 = 71.3^{+4.9}_{-4.5} \mathrm{~km} \mathrm{~s}^{-1} \mathrm{Mpc}^{-1}$. Using SLACS IX velocity dispersion alone gives an $H_0 = 68.2^{+4.4}_{-4.5} \mathrm{~km} \mathrm{~s}^{-1} \mathrm{Mpc}^{-1}$. After fully accounting for the selection functions, velocity dispersion bias, and implementing this in the \citetalias{Birrer2020} hierarchical inference code, we find that the Hubble constant becomes: $H_0 = 65.6^{+4.4}_{-4.4} \mathrm{~km} \mathrm{~s}^{-1} \mathrm{Mpc}^{-1}$

All flexible lensing plus dynamics approaches to breaking the mass sheet degeneracy are fundamentally limited by our ability to measure velocity dispersions precisely. Before \citet{Knabeletal2024} reported a 5.3$\%$ bias on velocity dispersions, we were about to publish $H_0 = 63.6^{+4.8}_{-3.9} \mathrm{~km} \mathrm{~s}^{-1} \mathrm{Mpc}^{-1}$, based on the SDSS fibre velocity dispersions used in \citetalias{Birrer2020}. Since  \citet{Knabeletal2024} does not have KCWI velocity dispersions for each of the 33 SLACS lenses used in \citetalias{Birrer2020}, we ultimately decided to use the \citet{slacs9} velocity dispersions which are available for all 33 lenses and are broadly consistent with \citet{Knabeletal2024} where there is overlap. For the non-overlapping sample \citet{Bolton:2008xf} and \citet{slacs9} more closely agree than for the lenses with KCWI data in \citet{Knabeletal2024}: had they all been 5.3$\%$ higher we would have found  $H_0 = 68.2^{+4.4}_{-4.5} \mathrm{~km} \mathrm{~s}^{-1} \mathrm{Mpc}^{-1}$. 

We investigated several potential sources of systematics that could be introduced by changing our parent deflector population. The average $\lambda_\mathrm{int}$ shifts is around 2-6$\%$ depending on the galaxy model we use to construct the catalog. The shifts in the inferred $H_0$ in the context of the current uncertainties of the measurement are modest. This indicates that our result is robust to the astrophysical uncertainties we have looked at. 

Putting together the statistical uncertainties and the systematics we have investigated, we find $H_0 = 66\pm4 \mathrm{(stat)} \pm 1 \mathrm{(model \: sys)} \pm 2 \mathrm{(measurement \: sys)} \mathrm{~km} \mathrm{~s}^{-1} \mathrm{Mpc}^{-1}$ for the TDCOSMO plus SLACS dataset (shown in the figure \ref{fig:h02}). The first systematic error accounts for the plausible alternative choices investigated in Section \ref{sec:discussion}, and the second is due to systematic errors in the measurement of velocity dispersions for SLACS lenses between \citet{Bolton:2008xf}, \citet{slacs9} and \citet{Knabeletal2024}.

Whilst our final result of $H_0 = 66\pm4 \mathrm{(stat)} \pm 1 \mathrm{(model \: sys)} \pm 2 \mathrm{(measurement \: sys)} \mathrm{~km} \mathrm{~s}^{-1} \mathrm{Mpc}^{-1}$ for the TDCOSMO plus SLACS dataset (shown in the figure \ref{fig:h02}) is apparently close to the $H_0=67.4_{-3.2}^{+4.1} \mathrm{~km} \mathrm{~s}^{-1} \mathrm{Mpc}^{-1}$ reported in \citetalias{Birrer2020}, it is important to remember that we have accounted for two effects that happened approximately cancel out. The selection function lowered $H_0$ by $\sim$3\% and the change to SLACS IX almost completely undid this shift. Had we made only one of the changes the results would differ by about 1 sigma.

In future large samples, we should expect to see similar offsets in the population level differences between galaxy-galaxy lenses and quadruple image quasars, even with lenses discovered in imaging surveys and fainter lensed quasar populations. The galaxy model in our simulation should be more complex to include more potential systematic effects for future large samples.

Twinning the observed properties (e.g. $R_\mathrm{eff}/\theta_E$) of the galaxy-galaxy and galaxy-quasar populations cannot remove the differences in the underlying mean population $\lambda_\mathrm{int}$ values if there is a strong selection function between the time-delay lenses and galaxy-galaxy lenses. This work has shown that in order to model their selection functions, we need to have a better understanding of the galaxy mass profile (e.g. dark matter inner slope).
This work has shown that using galaxy-galaxy lenses to break the mass-sheet degeneracy for galaxy-quasar lens populations requires an understanding of population selection functions. At the moment, different reasonable choices for modelling this selection function place a $\sim$  $\mathrm{~km} \mathrm{~s}^{-1} \mathrm{Mpc}^{-1}$ floor on how well the mass sheet degeneracy can be broken for time delay cosmology.

\section*{Software and Data Availability}
The software used in this work for the simulation and analysis of the lens sample is  lenstronomy\footnote{\url{https://github.com/lenstronomy/}} \citep{Birrer:2018xgm, Birrer:2015rpa}. Dark matter halo properties are calculated through COLOSSUS \footnote{\url{bitbucket.org/bdiemer/colossus}} \citep{2018ApJS..239...35D}. The the stellar mass was obtained by CosmoDC2 catalog  \footnote{\url{https://irsa.ipac.caltech.edu/Missions/cosmodc2.html}}

\section*{Acknowledgements}
We are particularly grateful to Simon Birrer for extensive discussions, comments on the manuscript and help with the use of the \citetalias{Birrer2020} notebooks.

We are grateful to Sherry Suyu, Anowar Shajib, Narayan Khadka, Alessandro Sonnenfeld, and Xiangyu Huang for helpful conversations that have enriched this work.

Numerical computations were done on the Sciama High Performance Compute (HPC) cluster which is supported by the ICG, SEPNet, and the University of Portsmouth. We are grateful to Coleman Krawczyk for help running our code on Sciama. 

This work has received funding from the European Research Council (ERC) under the European Union's Horizon 2020 research and innovation programme (LensEra: grant agreement No 945536). TEC is funded by the Royal Society through a University Research Fellowship. The work of PJM and SME was supported by the U.S. Department of Energy under
contract number DE-AC02-76SF00515.
For the purpose of open access, the authors have applied a Creative Commons Attribution (CC BY) license to any Author Accepted Manuscript version arising.


\bibliographystyle{mnras}
\bibliography{H0Selection} 



\appendix
\onecolumn
\section{Population}

\begin{figure}
    \centering
    \includegraphics[width=0.8\textwidth]{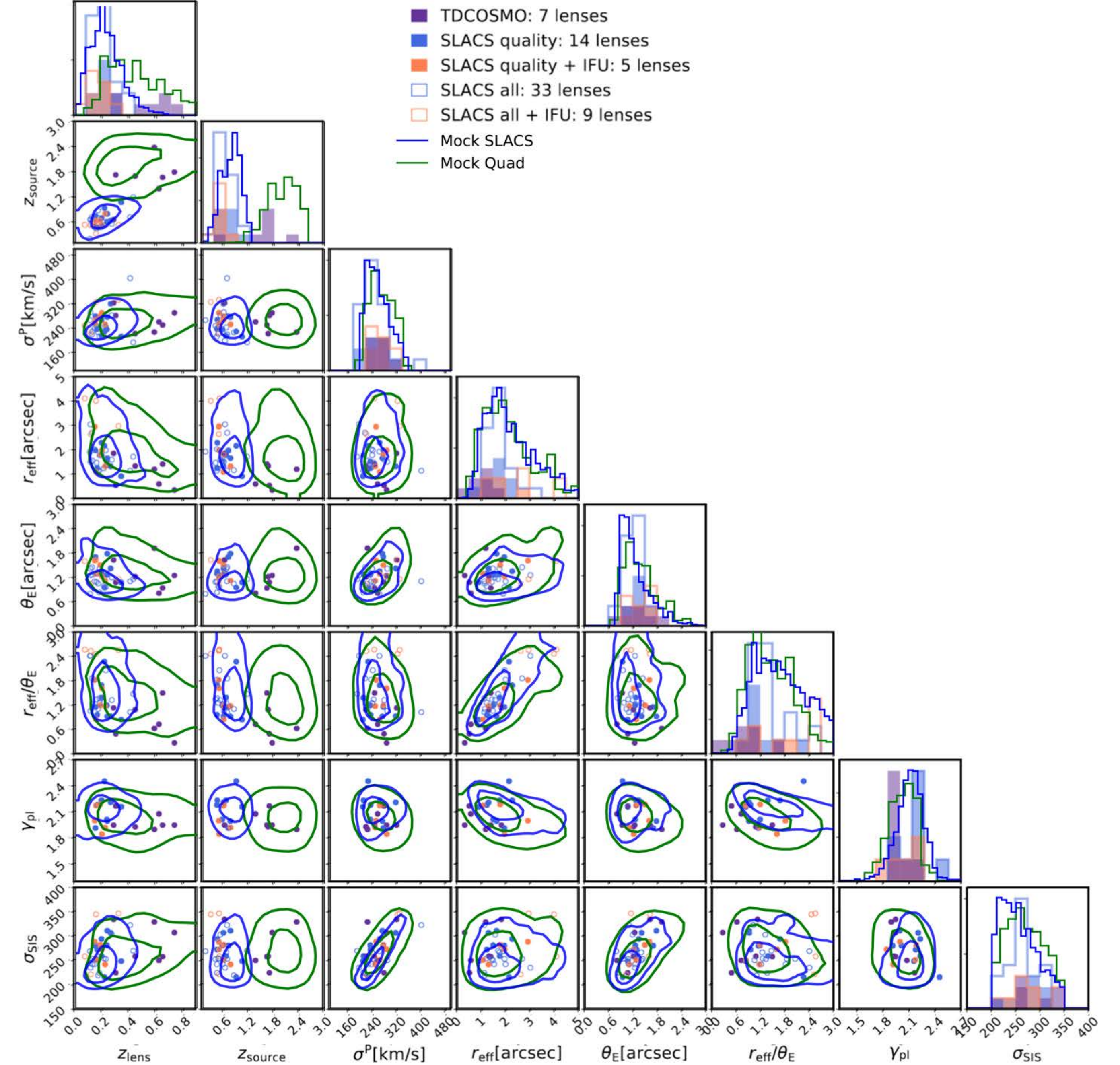}
    \caption{Distribution of the parameters of the mock population from our fiducial model compare to \citetalias{Birrer2020}. The blue and green contours are the Mock SLACS lenses and MOCK Lensed Quasars, respectively}
    \label{fig:population}
\end{figure}

Figure \ref{fig:population} shows the distribution of the parameters of the mock population from our fiducial model overlapping with Figure 8 in \citetalias{Birrer2020} before twining $R_\mathrm{eff}/\theta_\mathrm{E}$. In general, our fiducial model is able to reflect the actual distribution of the SLACS galaxy-galaxy lenses and TDCOSMO lensed quasars.

\section{Constant TSHMR}
\begin{figure*}
    \label{app:constantMh}
    \centering
    \includegraphics[width = \textwidth]{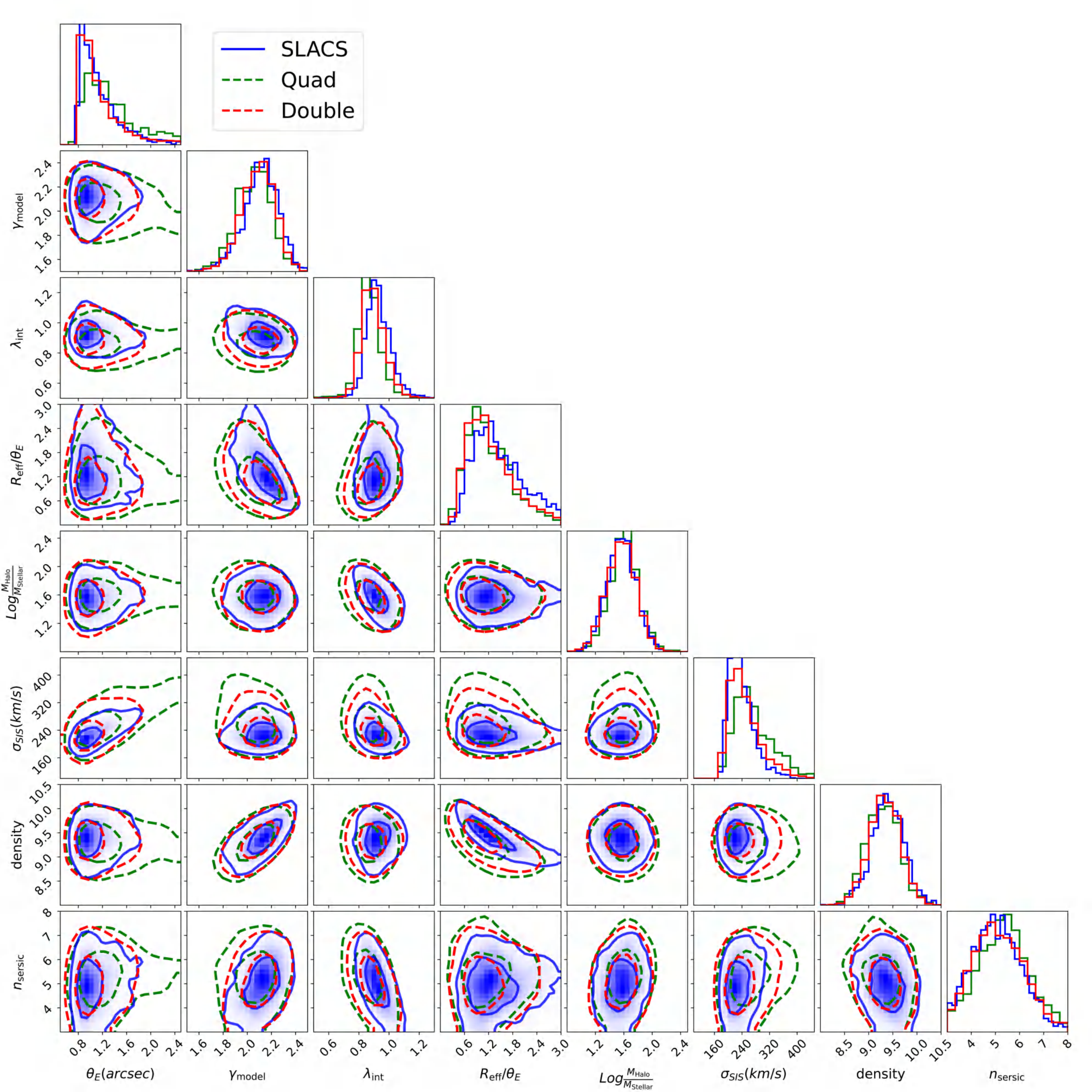}
    \begin{tabular}{ccccccccccc} 
    \hline
    Name & $\theta_E ('')$ & $\gamma_{\text{local}}$ & $\gamma_{\mathrm{model}}$ & $\lambda_\mathrm{int}$ & $R_\mathrm{eff}(kpc)$ & $M_\star (M_\odot)$ & $M_{DM}(M_\odot)$ & $log\frac{M_\mathrm{DM}}{M_\star}$ & $\sigma_{SIS} (km/s)$ & $n_{\text{sersic}}$\\
    \hline
    Double & $0.73^{+0.37}_{-0.16}$ & $1.98^{+0.12}_{-0.12}$ & $2.10^{+0.14}_{-0.13}$ & $0.90^{+0.06}_{-0.06}$ & $5.8^{+3.8}_{-2.1}$ & $11.3^{+0.3}_{-0.2}$ & $12.9^{+0.3}_{-0.3}$ & $1.55^{+0.21}_{-0.20}$ & $212^{+47}_{-29}$ & $4.91^{+0.79}_{-0.75}$\\
    Quad & $1.25^{+0.68}_{-0.30}$ & $1.90^{+0.13}_{-0.11}$ & $2.05^{+0.14}_{-0.13}$ & $0.87^{+0.06}_{-0.05}$ & $9.2^{+7.3}_{-3.8}$ & $11.7^{+0.3}_{-0.3}$ & $13.3^{+0.3}_{-0.3}$ & $1.60^{+0.19}_{-0.21}$ & $262^{+69}_{-43}$ & $5.24^{+0.81}_{-0.64}$\\
    SLACS & $1.05^{+0.37}_{-0.19}$ & $2.04^{+0.12}_{-0.13}$ & $2.12^{+0.14}_{-0.13}$ & $0.92^{+0.07}_{-0.06}$ & $6.0^{+4.7}_{-2.5}$ & $11.5^{+0.3}_{-0.2}$ & $13.0^{+0.3}_{-0.3}$ & $1.57^{+0.20}_{-0.21}$ & $229^{+38}_{-25}$ & $4.98^{+0.75}_{-0.76}$\\
    \hline
    \end{tabular}
    \caption{The distribution of SLACS lenses (blue), monitorable quads (green), and monitorable double (red). SLACS samples are selected that the Einstein radius has to be larger than 0.8$''$ and must be modelable. The bottom table shows the corresponding distribution.}
    \label{tab:Constant}
\end{figure*}
In this appendix We also used a difference choice in \cite{Sonnenfeld2023}, where the halo masses are drawn  from the following distribution:
\begin{equation}
\label{eq:constanttshmr}
\log M_{\mathrm{halo}}\ \sim  \log M_\star + \mathcal{N}\left(1.5, 0.2\right),
\end{equation}
This relation comes from weak lensing constraints on the halo mass of elliptical galaxies \citep{Sonnenfeld2022}. The difference between the two TSHMR is that the log $\frac{M_\mathrm{halo}}{M_\star}$ is proportional to stellar mass in equation \ref{eq:huangtshmr}, and the mean value of this ratio is higher. The results are shown in figure \ref{app:constantMh} and discussed in section \ref{sec:tshmr}.


\bsp	
\label{lastpage}
\end{document}